\documentclass[10pt,english,twocolumn,nofootinbib]{revtex4-2}

\usepackage[english]{babel}
\usepackage[T1]{fontenc}
\usepackage[utf8]{inputenc} 
\usepackage[version=4]{mhchem}
\usepackage{geometry, units, textcomp, amsmath, graphicx, xcolor, lipsum, xfrac, comment, upgreek, appendix}
\renewcommand{\thefootnote}{\fnsymbol{footnote}}

\begin{document}

\author{J. Travesedo$^{1\dagger}$, J. O'Sullivan$^{1\dagger}$, L. Pallegoix$^1$, Z. W. Huang$^1$, P. Hogan$^2$, P. Goldner$^3$, T. Chaneliere$^4$, S. Bertaina$^5$, D. Est\`eve$^1$, P. Abgrall$^1$, D. Vion$^1$, E. Flurin$^1$, P. Bertet$^{1}$}

\email{patrice.bertet@cea.fr}

\affiliation{$^1$Universit\'e Paris-Saclay, CEA, CNRS, SPEC, 91191 Gif-sur-Yvette Cedex, France\\
$^2$University College London, LCN, QSD \\ 17-19 Gordon Street, London WC1H 0AH, United Kingdom\\$^3$Chimie ParisTech, PSL University, CNRS, Institut de Recherche de Chimie Paris, 75005 Paris, France\\$^4$Universit\'e Grenoble Alpes, CNRS, Grenoble, France \\
$^5$Aix-Marseille Univ. University of Toulon, IM2NP, 13013, Marseille, France}


\title{All-microwave spectroscopy and polarization of individual nuclear spins in a solid}

\begin{abstract}

\textbf{
We report magnetic resonance spectroscopy measurements of individual nuclear spins in a crystal coupled to a neighbouring paramagnetic center, detected using microwave fluorescence at millikelvin temperatures. We observe real-time quantum jumps of the nuclear spin state, a proof of their individual nature. By driving the forbidden transitions of the coupled electron-nuclear spin system, we also achieve single-spin solid-effect dynamical nuclear polarization. Relying exclusively on microwave driving and microwave photon counting, the methods reported here are in principle applicable to a large number of electron-nuclear spin systems, in a wide variety of samples.
} 
\end{abstract}

\maketitle
\def\thefootnote{$\dagger$}\footnotetext{Both authors contributed equally to this work}\def\thefootnote{\arabic{footnote}}

Pushing the sensitivity of nuclear magnetic resonance spectroscopy to the single spin level would have a major impact in chemistry and biology and is the goal of intense research efforts ~\cite{budakian_roadmap_2024,grob_magnetic_2019,mamin_nanoscale_2013}. However, measuring the magnetic resonance spectrum of individual nuclei is made difficult by the small value of the nuclear magnetic moment. The sensitivity to directly detect as few as $\sim 100$ nuclear spins has been achieved using mechanical~\cite{mamin_isotope-selective_2009,grob_magnetic_2019,budakian_roadmap_2024} and atomic spin~\cite{mamin_nanoscale_2013,staudacher_nuclear_2013,du_single-molecule_2024} sensors. So far however, it has only been possible to measure individual nuclear spins through their hyperfine coupling to an individual paramagnetic electronic system, whose spin then needs to be detected. Certain electron spin systems can be detected optically using spin-dependent photoluminescence. This has enabled optical detection of individual $^{13}\mathrm{C}$ nuclear spins by NV centers in diamond~\cite{neumann_single-shot_2010,taminiau_detection_2012,cujia_tracking_2019}, of individual $^{73}$Ge nuclei by a GeV center in diamond~\cite{adambukulam_hyperfine_2024}, of individual $^{13}\mathrm{C}$ and $^{29}\mathrm{Si}$ by di-vacancy defects in silicon carbide~\cite{bourassa_entanglement_2020}, and of individual $^{1}\mathrm{H}$~\cite{uysal_coherent_2023} and $^{29}\mathrm{Si}$~\cite{kornher_sensing_2020} nuclei by rare-earth-ions in oxide crystals. In other systems, the electron spin can be converted into a charge, which can be detected using electrical currents. This has enabled electrical detection of individual $^{31}$P~\cite{pla_high-fidelity_2013}, $^{29}$Si~\cite{pla_coherent_2014}, and $^{123}$Sb~\cite{asaad_coherent_2020} nuclear spins by a phosphorus donor in silicon, and of individual $^{159}$Tb nuclear spins in a $\mathrm{TbPc}_2$ single-molecule magnet~\cite{vincent_electronic_2012}. Electrical detection by a scanning tunnelling microscope of the electron spin of individual atoms on a conducting surface enabled hyperfine spectroscopy of $^{57}$Fe, $^{47}$Ti, and $^{49}$Ti nuclear spins~\cite{willke_hyperfine_2018}, and spectroscopy and polarization of $^{63}$Cu and $^{65}$Cu~\cite{yang_electrically_2018} nuclear spins.


Recently, individual electron spins were detected using microwave photon counting at 10\,mK~\cite{wang_single-electron_2023}. This fluorescence detection method ~\cite{albertinale_detecting_2021,billaud_microwave_2023} relies solely on the magnetic coupling of the electron spin to a detection microwave resonator, whose role is to enhance the radiative spin relaxation rate $\Gamma_R$ via the Purcell effect~\cite{purcell_spontaneous_1946,bienfait_controlling_2016}. It is therefore in principle applicable to a larger class of paramagnetic systems than optical or electrical detection. Here, we show that fluorescence detection of a single electron spin also enables readout, spectroscopy, and polarization of individual nuclear spins of the host crystal to which the electronic probe is strongly coupled.

\begin{figure}[tbh!]
    \includegraphics[width=\columnwidth]{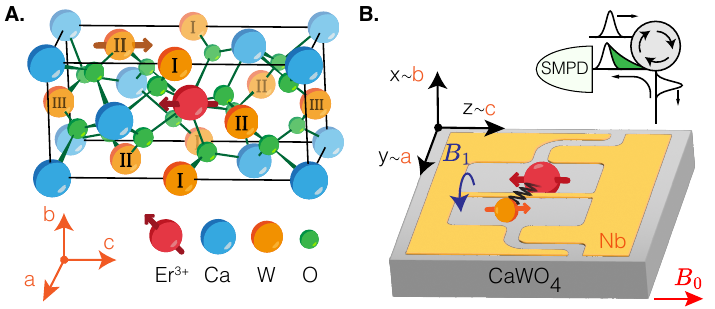}
    \caption{\label{fig1}
    \textbf{CaWO$_4$ crystal structure and experiment schematic}
    \textbf{A.} Schematic representation of the first unit cell of the CaWO$_4$ crystal, which has tetragonal symmetry around the $c$ axis. The crystalline axes \textit{(a,b,c)} are shown in orange. An Er$^{3+}$ ion (red) substitutes a calcium ion (blue). Three types of W sites (labeled as I, II, III) are distinguished, based on their location with respect to the Er$^{3+}$.
    \textbf{B.} Sample schematics. A resonator (yellow) is fabricated out of a niobium thin-film, on the surface of a CaWO$_4$ slab (grey). The nanowire constriction generates an oscillating magnetic field $B_1$, which couples to an Er$^{3+}$ electron spin in the vicinity (red). The latter may be coupled to one or several neighbouring $^{183}$W nuclear spin (orange). The sample is connected to the microwave lines through an antenna. A circulator routes coherent excitation pulses (Gaussian curves) to the sample and the fluorescence signal (green exponential decay) to an SMPD. The axes $(x, y, z)$ along the edges of the slab approximately match the crystalline axes $(a,b,c)$. A magnetic field $B_0$ is applied in the $(y,z)$ plane, approximately parallel to the wire.
    }
\end{figure}

\begin{figure*}[tbh!]
    \includegraphics[width=\textwidth]{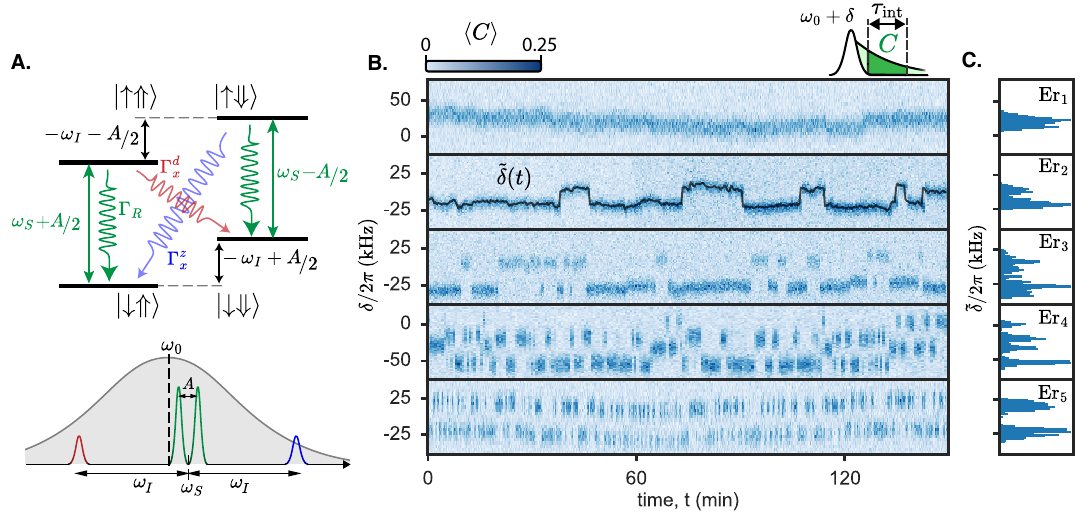}
    \caption{\label{fig2}
    \textbf{Single-$^{183}$W-nuclear-spin quantum jumps.}.
    \textbf{A.} \textit{top} Energy level diagram of the $^{183}$W-Er$^{3+}$ coupled system. Allowed electronic transitions are shown as straight green arrows. Relaxation rates are shown as wiggly arrows, green for allowed transitions (rate $\Gamma_R$), red for the double-quantum (rate $\Gamma_x^d$), and blue for the zero-quantum (rate $\Gamma_x^z$).
    \textit{Bottom}. Schematic of the $^{183}$W-Er$^{3+}$ transitions, superimposed with the resonator response shown as a gray Lorentzian of FWHM $\kappa$. Allowed transitions (green peaks) are quasi-resonant with $\omega_0$, forbidden transitions (red and blue) are detuned by $\sim \omega_I$. 
    \textbf{B.} \textit{Top.} Fluorescence detection spectroscopy pulse scheme. A Gaussian $\pi$-pulse of 80 \textmu s of duration and frequency $\omega_0$ + $\delta$ is applied to the sample. Ensemble-averaged number of counts $\langle C \rangle$ is subsequently measured during an integration time window, $\tau_{\mathrm{int}}$. The sequence is then repeated without any delay, with the excitation pulse frequency updated.
    \textit{bottom}. High-resolution spectra of five $\mathrm{Er}^{3+}$ ion spins, measured consecutively during $150$\,minutes. Each spectrum is recorded by averaging $200$ sweeps of $\delta$, yielding the ensemble-averaged number of counts $\langle C \rangle (\delta)$. The integration time $\tau_{\mathrm{int}}$ is 1.6 ms for $\mathrm{Er}_1$, 2.0 ms for $\mathrm{Er}_{2,3}$ and 3.6 ms for $\mathrm{Er}_{4,5}$. The increase in counts corresponds to the detected fluorescence after exciting an EPR-allowed transition. The resonance frequency evolves in time with a slow drift and sudden telegraphic jumps. Each spectrum is fitted as a Lorentzian and the center $\tilde{\delta}$ is shown as a function of time for $\mathrm{Er}_2$ as a black continuous line. 
    \textbf{C.} Histograms of $\tilde{\delta}$ for the five ions.
    }
\end{figure*}

\subsection*{Experimental setup and spin system}
For this demonstration we use a crystal of CaWO$_4$ (see Fig.~\ref{fig1}A). The nuclear spin species with the largest concentration in the crystal is $^{183}$W, a 14.4 \% abundant isotope with a nuclear spin $I=1/2$ and a low gyromagnetic ratio $\gamma_W/2\pi = 1.774$\,MHz/T~\cite{knight_solid-state_1986}. Individual $^{183}$W nuclear spins are detected by their hyperfine interaction with Er$^{3+}$ ions randomly located throughout the crystal. The lowest-energy Kramers doublet of Er$^{3+}$ ions form an effective electron spin $S=1/2$ with a gyromagnetic tensor $\bar{\bar{\gamma}}$ (see App.\ref{section:sample}) when they enter CaWO$_4$ in substitution to $\mathrm{Ca}^{2+}$~\cite{bertaina_rare-earth_2007,le_dantec_twenty-three-millisecond_nodate}. A microwave resonator of frequency $\omega_0/2\pi = 7.7492$\,GHz and linewidth $\kappa/2\pi = 640\,\mathrm{kHz}$ is patterned directly on top of the CaWO$_4$ sample (see Fig.~\ref{fig1}B). The resonator is made out of a superconducting thin-film of niobium and contains a narrow constriction around which the magnetic field $B_1$ is strong, enabling fluorescence detection of the nearby Er$^{3+}$ ions (see App.\ref{section:sample}). The resonator output is directed towards a Single Microwave Photon Detector (SMPD) based on a superconducting transmon qubit \cite{lescanne_irreversible_2020,albertinale_detecting_2021,wang_single-electron_2023,balembois_cyclically_2024}. For fluorescence detection, a microwave pulse of frequency $\omega_0$ is applied to the sample, and the number of counts $C$ is then recorded during an integration time window. A magnetic field $\bar{B}_0$ (amplitude $B_0$) is applied parallel to the sample surface, approximately along the crystalline $c$-axis. When the frequency $\omega_S = | \bar{\bar{\gamma}} \cdot \bar{B}_0 | $ of an Er$^{3+}$ ion is resonant with $\omega_0$, a peak in the ensemble-averaged $\langle C \rangle (B_0)$ is visible if the ion radiative relaxation rate $\Gamma_R$ is large enough (typically, larger than $10^3\,\mathrm{s}^{-1}$ In our device this condition is met for ions at a distance $\sim 150 \mathrm{nm}$ or less from the resonator constriction \cite{wang_single-electron_2023} (see App.\ref{section:sample}).

The interaction between an Er$^{3+}$ electron spin and a $^{183}\mathrm{W}$ nuclear spin can be described in the secular approximation by the Hamiltonian 

\begin{equation}
    \hat{H} / \hbar = \omega_S \hat{S}_z + \omega_I \hat{I}_z + A  \hat{S}_z \hat{I}_z  + B \hat{S}_z \hat{I}_x,
\end{equation}

where $\omega_I = -\gamma_{W} B_0$ is the $^{183}\mathrm{W}$ Larmor frequency, $A$ is the isotropic hyperfine coupling arising from both dipolar magnetic and contact interactions, and $B$ is the anisotropic hyperfine coupling whose origin is purely dipolar magnetic. Both $A$ and $B$ depend on the relative distance between the Er$^{3+}$ and the $^{183}\mathrm{W}$ as well as their relative orientation with respect to the magnetic field $\bar{B}_0$. Within the first unit cell, and with $\bar{B}_0$ applied parallel to the $c$-axis, three sets of magnetically-equivalent $^{183}\mathrm{W}$ atoms can be found, called Types I, II, and III in the following (see Fig. \ref{fig1}A). All $^{183}\mathrm{W}$ nuclear spins beyond the first unit cell have lower couplings and are not resolved in this study.


In the high-field limit $|\omega_I| \gg A, B$, the eigenstates of the two-spin system are close to the uncoupled spin states $|\downarrow \Uparrow\rangle$, $|\downarrow \Downarrow\rangle$, $|\uparrow \Uparrow\rangle$, $|\uparrow \Downarrow\rangle$ in ascending order of energy, as shown schematically in Fig. \ref{fig2}A. Because $\langle \downarrow \Uparrow | S_x | \uparrow \Uparrow \rangle \approx \langle \downarrow \Downarrow | S_x | \uparrow \Downarrow \rangle \approx 1/2$, the two nuclear-spin preserving transitions at $\sim \omega_S \mp A/2$ are EPR-allowed. They can be driven by a resonant microwave pulse; moreover, cavity-enhanced radiative relaxation at rate $\Gamma_R$ can take place on either transition, giving rise to the fluorescence detection signal. On the other hand, $ | \langle \downarrow \Downarrow | S_x | \uparrow \Uparrow \rangle | \approx |\langle \downarrow \Uparrow | S_x | \uparrow \Downarrow \rangle | \approx |B/(4\omega_I)|$, implying that the nuclear-spin flipping transition at $\omega_S + \omega_I $ (respectively, $\omega_S - \omega_I $) is not completely forbidden when $B$ is non-zero~\cite{schweiger_principles_2001}. This double-quantum (respectively, zero-quantum) forbidden transition can thus be microwave-driven. Moreover, radiative electron relaxation with nuclear-spin-flipping (called cross-relaxation in the following) also becomes weakly authorized, at a rate $\Gamma_x^{d,z} = \Gamma_R \frac{B^2}{4\omega_I^2 } \frac{1}{1 + 4[\omega_I \pm (\omega_S - \omega_0)  ]^2/\kappa^2}$. The last term is due to the off-resonant Purcell effect from the superconducting resonator. We note that the two cross-relaxation rates are equal only if the spin and cavity are on resonance ($\omega_S = \omega_0$). We also introduce the cross-relaxation probability $\eta^{d,z} \equiv \Gamma_x^{d,z} / (\Gamma_R + \Gamma_x^{d,z}) \approx \Gamma_x^{d,z} / \Gamma_R$.

\begin{figure}[t!]
    \includegraphics[width=\columnwidth]{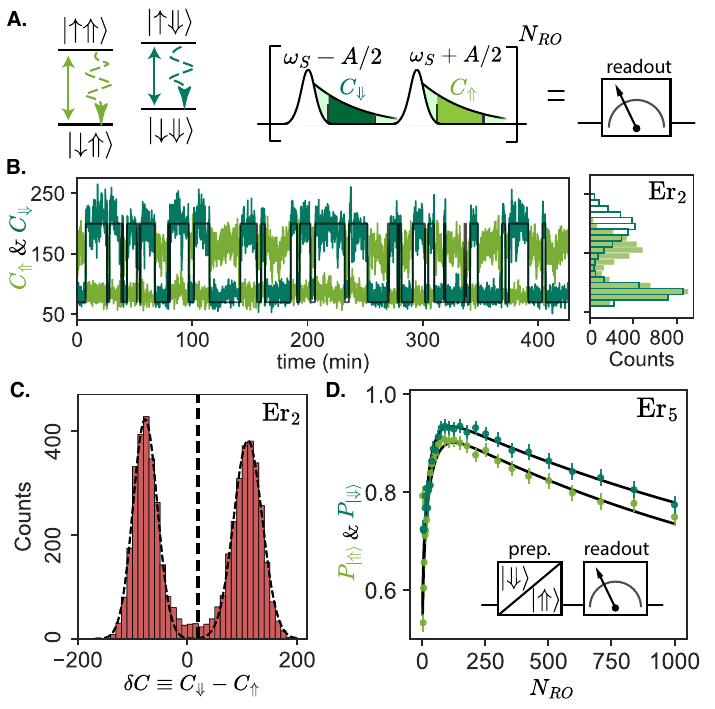}
    \caption{\label{fig3}
    \textbf{Single-shot $^{183}$W nuclear spin read-out}.
    \textbf{A.} \textit{left}. Energy level diagram of an $^{183}$W-Er$^{3+}$ coupled system. EPR-allowed trasitions are represented as solid bright and dark green lines and the respective radiative decays in dashed lines of the same color. \textit{right} Single-shot nuclear read-out pulse sequence. Resonant pulses are applied on each EPR-allowed transition and the number of counts are recorded independently for the two frequencies. The sequence is repeated $N_{RO}$ times, yielding the number of counts after each sequence, $C_\Downarrow$ and $C_\Uparrow$.
    \textbf{B.} \textit{Left} Integrated counts $C_\Downarrow$ (resp. $C_\Uparrow$) as a function of time as a dark green (resp. light green) solid line. The data was taken for $\mathrm{Er_2}$ with integration time $\tau_{\mathrm{int}}$ = 1.6 ms and $N_{RO}$ = 10$^3$. Quantum jumps appear as abrupt changes in the number of counts with telegraphic trajectory. The two traces are anti-correlated and the state of the nuclear spin can be mapped to $|\Downarrow\rangle$ when C$_\Downarrow$>C$_\Uparrow$, represented by a black continuous line. \textit{Right} Read-Histrograms of C$_\Downarrow$ and C$_\Uparrow$. The lower number number of counts in state $\Uparrow$ is attributed to imperfect SMPD frequency tuning, resulting in different fluorescence intensities. 
    \textbf{C.} Histogram of $\delta C$ $\equiv$ $C_\Downarrow$ - $C_\Uparrow$. A bimodal Gaussian distribution fit is plotted as black dashed line. Vertical black dashed line shows the threshold for single-shot read-out.
    \textbf{D.} Probability of measuring state $|\Downarrow\rangle$ and $|\Downarrow\rangle$ (resp. dark and light green) as a function of $N_{RO}$, after preparation in the respective state. Experimental data are shown as green dots and the fit is shown as a solid black line (see App.\ref{section:readout}). The data were obtained for $\mathrm{Er}_5$ with integration time $\tau_{\mathrm{int}}$ = 2 ms. \textit{Inset.} Pulse sequence. The nuclear spin is initialized into state $\Downarrow$ (resp. $\Uparrow$) followed by state readout sweeping $N_{RO}$.
    }
\end{figure}

\begin{figure*}[tbh!]
    \includegraphics[width=\textwidth]{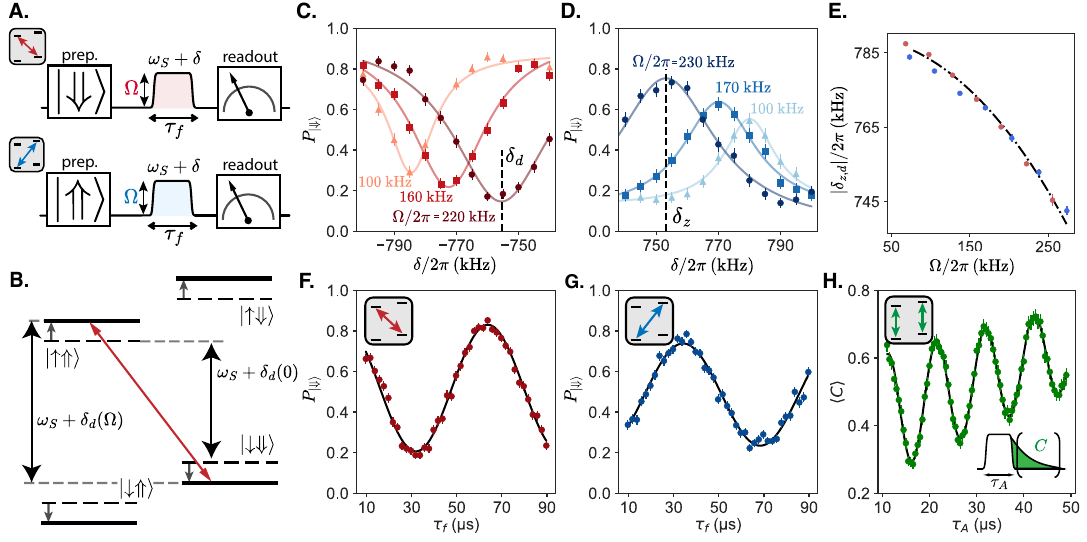}
    \caption{\label{fig4}
    \textbf{Single-spin ELDOR-detected NMR spectroscopy.}
    \textbf{A.} Pulse sequence. To measure the double-quantum (resp., zero-quantum) transition, the nuclear spin is first prepared in the $\Downarrow$ (resp., $\Uparrow$) state. A microwave pulse with a flattop-Gaussian envelope (maximum amplitude $\Omega$) and a frequency $\omega_S + \delta$ is applied, followed by nuclear-spin read-out.
    \textbf{B.} AC Zeeman shift diagram. The spin levels are frequency-shifted by the microwave drive, from their $\Omega=0$ value (black dashed lines) to the value under drive (solid black line).
    \textbf{C \& D}. Measured (dots) nuclear spin $\Downarrow$ probability as a function of $\delta$ for three values of $\Omega$ with pulse duration $\tau_f$ equal to 60 $\upmu$s, 32 $\upmu$s and 22 $\upmu$s in increasing order of $\Omega$. Panel $c$ (resp. $d$) shows the double-quantum (resp. zero-quantum) transition. Solid lines are Lorentzian fits, yielding the measured transition frequency $\delta_d$ (resp. $\delta_z$) 
    \textbf{e.} Measured double-quantum (resp. zero-quantum) frequency $|\delta_d|$ (resp. $|\delta_z|$) as a function of $\Omega$. Dash-dotted line is a fit, using the AC Zeeman shift model described in App.\ref{section:omega_I}.
    \textbf{F \& G}. Measured nuclear spin $\Downarrow$ probability as a function of $\tau_{f}$ for the double-quantum (f, red dots) and zero-quantum (g, blue dots) sequences, with $\delta$ set at resonance, for $\Omega/2\pi = 220$ kHz. Black solid lines are cosine fits, yielding a forbidden Rabi frequency $\Omega_{z/d}/2\pi \:\sim$ 15 kHz for both of them. \textit{Insets.} Diagram representing which transition is being driven.
    \textbf{H.} Allowed transition Rabi oscillation. A pulse at $\omega_S$ of duration $\tau_A$ is applied followed by fluorescence detection. The input pulse amplitude is $6.2$ times lower than in panels f and g. Green dots are ensemble-averaged number of counts $\langle C \rangle$ as a function of $\tau_A$. Solid black line is a cosine fit with linearly increasing offset, yielding $\Omega/2\pi = 97$ kHz. The Rabi frequency is not 6.2 times smaller due to the difference in resonator filtering. \textit{Inset.} \textit{Top}. Transition diagram.
    All data in this figure are from $\mathrm{Er_5}$.
    }
\end{figure*}

\subsection*{Nuclear spin quantum jumps}
We identify individual $\mathrm{Er}^{3+}$ ion spins with $\Gamma_R \sim 10^3 \:\mathrm{s}^{-1} $ by scanning the field amplitude $B_0$ in a region around $446$\,mT (close to the $\mathrm{Er}^{3+}$ $I=0$ main resonance line), and identifying single peaks as explained in \cite{wang_single-electron_2023} (see App.\ref{section:electron}). We perform high-resolution spectroscopy by exciting the spin with an 80 \textmu s long Gaussian-shaped microwave $\pi$-pulse, with frequency $\omega$ followed by microwave fluorescence detection, sweeping $\delta = \omega - \omega_0$ over $2\pi \times 100$\,kHz. Each sweep is averaged 200 times to obtain the ensemble-averaged number of counts $\langle C \rangle (\omega)$. Figure ~\ref{fig2}B shows such spectra measured repeatedly during 150 minutes, for five representative ions labeled as $\mathrm{Er}_1 - \mathrm{Er}_5$. Ion $\mathrm{Er}_1$ displays a single line, whose center frequency fluctuates on a time scale of several minutes, and on a frequency scale of $\sim 20$\,kHz. This spectral diffusion is attributed to a combination of $B_0$ drift, charge noise, and re-arrangements of the weakly coupled nuclear spin bath, and was observed for all measured ions although with a different intensity (see discussion below). Because the drift is slow, it can be compensated by various frequency-tracking strategies (see App. \ref{section:tracking}) which are used in all the measurements reported later in this work. In contrast, the electron spin frequencies of ions $\mathrm{Er}_2$, $\mathrm{Er}_3$, and $\mathrm{Er}_5$ display a clear telegraphic noise between two frequencies, four in the case of $\mathrm{Er}_4$. As demonstrated in the following, these jumps occur when one nearby $^{183}\mathrm{W}$ nuclear spin changes state. Quantum jumps are a hallmark of the measurement of individual quantum systems, and have been observed for trapped ions~\cite{nagourney_shelved_1986,bergquist_observation_1986,sauter_observation_1986}, photons in a high-Q microwave cavity~\cite{gleyzes_quantum_2007,guerlin_progressive_2007}, superconducting circuits~\cite{vijay_observation_2011-1}, and various nuclear spin systems~\cite{neumann_single-shot_2010,pla_high-fidelity_2013,thiele_electrical_2013}; here, we report their observation on individual $^{183}\mathrm{W}$ nuclear spins by microwave fluorescence detection.

The transition frequency of each spectrum $\tilde{\delta}(t)$ is obtained from a Lorentzian fit, shown as a histogram in Fig.~\ref{fig2}C for the five ions. Each histogram presents a different number of resolved distributions. For $\mathrm{Er}_2$, $\mathrm{Er}_3$ and $\mathrm{Er}_5$, the two resolved distributions indicate the presence of a strongly coupled neighbouring nuclear spin. The size of the jumps is a direct measure of the hyperfine coupling constant $|A|$ (see App.\ref{section:traces}). In contrast $\mathrm{Er}_4$, shows four resolved distributions which correspond to two strongly coupled nuclear spins. The pair will be referred as $\mathrm{Er}_4^{(1)}$ and $\mathrm{Er}_4^{(2)}$ in the following. We note that the standard deviation of the individual distributions is significantly different for each ion, indicating that part of the drift originates from their local environment, likely from un-resolved nuclear spin jumps.

The direct relaxation rate of $^{183}$W nuclear spins proximal to Er$^{3+}$ impurities in CaWO$_4$ is exceedingly slow at 10\,mK, with an upper bound of $\sim 10^{-6}\,\mathrm{s}^{-1}$~\cite{wang_month-long-lifetime_2024}. We therefore attribute all observed quantum jumps to cross-relaxation events occurring via the Er$^{3+}$ spin excited state, enabling to determine cross-relaxation probabilities $\eta^{d,z}$ from each ion time trace (see Table 2 in App.\ref{section:traces}). A significant difference between $\eta^d$ and $\eta^z$ is observed for $\mathrm{Er}_{2,3,4}$, as also evidenced by the different peak weights in the histograms of $\tilde{\delta}$. The spin-resonator residual detuning (less than $30\,$kHz) is too low to explain this asymmetry in the radiative cross-relaxation model. This may indicate the existence of an extra, non-radiative, asymmetric cross-relaxation channel, possibly arising from the coupling to another detuned Er$^{3+}$ ion~\cite{wenckebach_dynamic_2019}. From the lowest value of $\eta^{d,z}$, we use the expressions above and extract $B$ for each nuclear spin (Shown, together with $A$, in Table I of App.\ref{section:dipole}). Based on calculated values of $A$ and $B$ assuming a purely dipolar interaction, we tentatively assign the nuclear spins $\mathrm{Er}_{2}$ and $\mathrm{Er}_{4}^{(2)}$ to Type III sites, the nuclear spin of $\mathrm{Er}_{5}$ to a Type I site, and the nuclear spin of $\mathrm{Er}_{3}$ and $\mathrm{Er}_{4}^{(1)}$ to either a Type I or a Type II site (see App.~\ref{section:dipole}).


\subsection*{Nuclear spin state readout}
We leverage the ability to spectrally resolve the different allowed transition frequencies of the electronic spin to perform a projective, quantum-non-demolition measurement of the nuclear spin state. We apply sequences of frequency-resolved excitation pulses (Gaussian-shaped $80 \:\mu\mathrm{s}$-long) at $\omega_S - A/2$ and $\omega_S + A/2$, spaced by an integration time $\tau_{\mathrm{int}}$ during which the fluorescence counts are measured; repeating this $N_{RO}$ times yields the number of counts after each sequence $C_\Uparrow$ and $C_\Downarrow$ (see Fig. \ref{fig3}A). 
Time traces over $7$ hours for these two quantities taken with ion $\mathrm{Er}_2$ are shown in Fig. \ref{fig3}B, using $N_{RO} = 10^3$ readout pulses. The two traces are clearly anti-correlated. The lower signal level in $C_\Uparrow(t)$ compared to $C_\Downarrow(t)$ is attributed to imperfect tuning of the SMPD center frequency. The photon-counting histograms for each time trace show two well separated distributions. 
From the histogram of $\delta C \equiv C_\Downarrow - C_\Uparrow$, which shows two separated and symmetric peaks, we define a threshold enabling single-shot readout of the nuclear spin state. Data shown in Fig \ref{fig3}C and D are for ion $\mathrm{Er}_2$. 


We prepare the nuclear spin in state $|\Downarrow\rangle$ (resp. $|\Uparrow\rangle$) (see below), and measure $P_{|\Downarrow\rangle}$ (resp. $P_{|\Uparrow\rangle}$) as a function of the number of readout pulses, $N_{RO}$, for ion $\mathrm{Er}_5$. At first, the probability of a correct read-out increases with $N_{RO}$, due to the increase in signal-to-noise ratio. After reaching a maximum value, $P_{|\Downarrow\rangle}$ then slowly decreases with $N_{RO}$, due to the finite cross-relaxation probability $\eta^{z, d}$ at each readout step. The analytical model used to describe the two mechanisms (see App.\ref{section:readout}) reproduces quantitatively the observations and yields an estimated value of $B/2\pi \sim 70\pm7$~kHz, close to the value extracted from the time traces. This indicates that cross-relaxation is the limiting factor for nuclear spin readout fidelity. It could be suppressed further in future experiments by increasing the resonator quality factor, hence the cavity filtering.


\subsection*{Single nuclear spin spectroscopy}

Nuclear spin spectroscopy is achieved by driving the forbidden transitions, in a single-spin version of the Electron-Double-Resonance-detected NMR (ELDOR-detected NMR) experiment ~\cite{schosseler_pulsed_1994}. Following preparation in state $|\Downarrow\rangle$ (resp. $|\Uparrow\rangle$) (see below) a flattop-shaped pulse is applied at a frequency $\omega_S + \delta$, The maximum amplitude of the pulse, $\Omega$, is quantified by the Rabi frequency of the allowed transitions (see Fig \ref{fig4}A and B). A reduction (resp. increase) in the nuclear spin state probability $P_\Downarrow$ is observed when $\omega_S + \delta$ matches the forbidden double-quantum (resp. zero-quantum) transition. Data are shown in Fig. \ref{fig4}C (resp. D) for $3$ values of $\Omega$, showing a frequency shift and power-broadening.


The fitted frequencies for the double- and zero-quantum transitions,  $\delta_d$ and $\delta_z$ respectively, are shown in Fig.~\ref{fig4}E as a function of $\Omega$. $|\delta_{z,d}|$ are found to depend quadratically on $\Omega$. This is due to off-resonant driving of the allowed transitions, which leads to AC-Zeeman shifts of the energy levels (see Fig \ref{fig4}B). In the high field limit, $|\delta_{z, d}| \approx - \omega_I - \sfrac{\Omega^2}{2|\delta_{d, z}|}$. The combined values of $|\delta_z|$ and $|\delta_d|$ are fitted to the complete analytical expression (see App.\ref{section:omega_I}), yielding the nuclear spin Larmor frequency, $\omega_I/2\pi = - 788.1\pm4$ kHz. The corresponding gyromagnetic factor of $1.767\pm1$~MHz\,$T^{-1}$ is close to the value expected for $^{183}\mathrm{W}$ in CaWO4, $\gamma_W/2\pi = 1.774$~MHz\,$T^{-1}$ \cite{knight_solid-state_1986}, thus confirming the nature of the nuclear spin. 

Driving the forbidden transitions provides an alternative method to determine $B$. Indeed, the zero-quantum (resp double-quantum) Rabi frequency $\Omega_{z}$ (resp. $\Omega_{d}$) is related to the allowed-transition Rabi frequency as $\Omega_{z/d} \approx \Omega \frac{B}{2\omega_I}$. Double- and zero-quantum Rabi oscillations are shown in Fig. \ref{fig4}F \& G, and allowed Rabi oscillations in  Fig.~\ref{fig4}H. 
The value obtained by this method, $|B|/2\pi=103\pm7$~kHz, is $\sim 30$~kHz larger than the value previously determined, possibly pointing to an underestimation of the uncertainties in the cross-relaxation method (see App. G). 



\begin{figure}[t!]
    \includegraphics[width=\columnwidth]{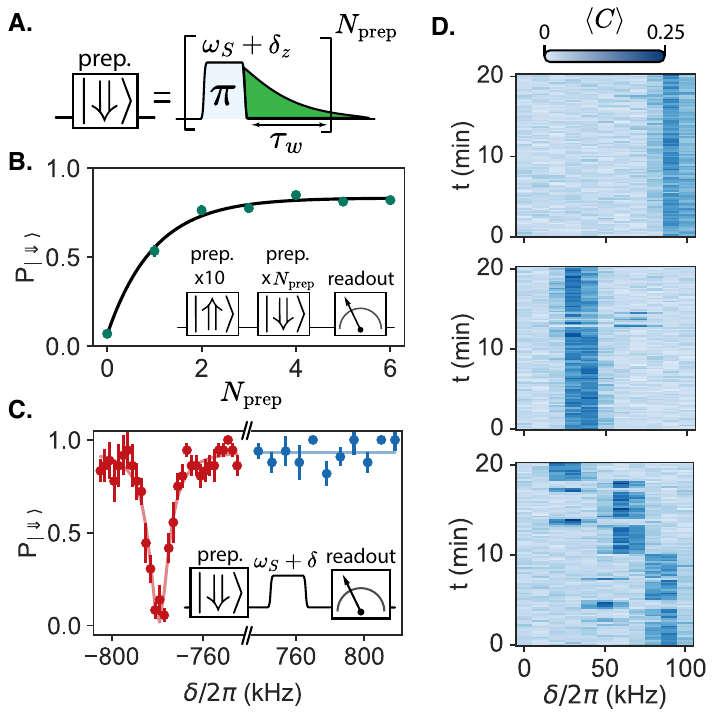}
    \caption{\label{fig5}
    \textbf{Single-spin dynamical nuclear polarization via solid effect.}
    \textbf{a.} Pulse sequence for preparing the nuclear spin in $\Downarrow$ (respectively, $\Uparrow$) through solid-effect DNP. One $\pi$-pulse at $\omega_S + \delta_z$ (respectively, $\omega_S + \delta_d$), followed by a relaxation time $\tau_w$, prepares the nuclear spin state to the desired state. The sequence is repeated $N_{\mathrm{prep}}$ times for higher efficiency. \textbf{b.}  Measured (dots) nuclear spin $\Downarrow$ probability as a function of the number of preparation pulses $N_{\mathrm{prep}}$ into this state. Solid line shows an exponential fit. \textit{Inset.} Pulse sequence diagram. The nuclear spin is initialized into the $\Uparrow$ state. Then, the preparation sequence for different $N_{\mathrm{prep}}$ and $\tau_w$ = 3 ms is applied. Finally, the state of the nuclear spin is measured.
    \textbf{c.} Probability of measuring the nuclear spin in state $\Downarrow$ in an ELDOR-detected NMR experiment for the zero- and double-quantum transitions after preparation in the $\Downarrow$ state. \textit{Inset.} ELDOR-detected NMR sequence. $N_{\mathrm{prep}}$ = 10 pulses and $\tau_w$ = 5 ms are used for preparation.
    \textbf{d.} Ensemble-averaged counts $\langle C \rangle$ measured in an fluorescence detection spectroscopy experiment as a function of frequency during 20 minutes. Data obtained for $\mathrm{Er_4}$ which has 2 neighbouring nuclear spins. From top to bottom, the nuclear spins are prepared to $|\Uparrow\Uparrow\rangle$, $|\Downarrow\Downarrow\rangle$ and not prepared. When preparing the state, the electron spin frequency $\tilde{\delta}$ remains constant (except three spectra where preparation was unsuccessful). In absence of state preparation, the frequency $\tilde{\delta}$ changes after each quantum jump.
    }
\end{figure}

\subsection*{Single nuclear spin polarization}
The nuclear spin state can be initialized via solid-effect Dynamic Nuclear Polarization (DNP) \cite{abragam_a_procto_w_nouvelle_1958,abragam_principles_1978,wenckebach_solid_2008}. To initialize the nuclear spin in $|\Uparrow\rangle$ (see Fig.~\ref{fig5}A), we apply a pulse on the zero-quantum transition, wait for a time $\tau_w$ (taken to be close to $\Gamma_R^{-1}$), and repeat $N_{\mathrm{prep}}$ times. $P_\Downarrow$ increases rapidly, reaching $\approx 0.8$ for $2$ pulses (Fig.~\ref{fig5}B). This demonstrates single-atom solid-effect nuclear spin polarization. The maximum value is limited by nuclear-spin read-out errors and out-of-equilibrium excitations of the electron spin, and can likely be improved in future experiments. 

State preparation is also confirmed spectroscopically (Fig \ref{fig5}C). After polarizing the state of the nuclear spin to $|\Downarrow\rangle$, a pulse is applied at a frequency $\omega_S + \delta$, with the amplitude corresponding to a $\pi$-pulse for the forbidden transition at resonance, followed by nuclear spin state readout. The probability $P_\Downarrow$ shows a clear dip at the frequency corresponding to the double-quantum resonance, but no dip on the zero-quantum transition, which is further evidence for nuclear spin polarization.

The solid-effect DNP is also effective when multiple spins are present, as seen in Fig.~\ref{fig5}D where it is applied to $\mathrm{Er}_4$.  
The polarization protocol into $|\Downarrow\Downarrow\rangle$, involves sending pulses at each of the zero-quantum transition frequencies. A $20$-minute time trace of the allowed-transitions spectrum is shown in Fig.~\ref{fig5}D in absence of (bottom), with polarization into $|\Downarrow\Downarrow\rangle$ (middle), and into $|\Uparrow \Uparrow\rangle$ (top). The polarization sequence is applied at the beginning of every spectrum. Whereas the trace with no polarization shows nuclear spin quantum jumps as reported in Fig.~\ref{fig2}B, the two other traces show dominantly the one allowed transition corresponding to the polarizated state. 

\subsection*{Discussion and outlook}

A logical next step will be to extend these methods to nuclear spins more weakly coupled to the electron spin, using methods demonstrated with NV centers in diamond~\cite{taminiau_detection_2012,cujia_tracking_2019,du_single-molecule_2024}. This would open the way to $^{183}\mathrm{W}$-nuclear-spin-based quantum registers operated and read-out via the $\mathrm{Er}^{3+}$ spin, generalizing recent $^{13}\mathrm{C}$ quantum registers demonstrations operated by a NV center in diamond~\cite{bradley_ten-qubit_2019,abobeih_fault-tolerant_2022}. Perhaps the most interesting aspect of the methods reported here is their applicability to a large range of nuclear spin systems in a large variety of samples, crystalline or not. Indeed, individual paramagnetic centers can be detected by microwave fluorescence provided their Purcell relaxation rate $\Gamma_R$ can be increased above $\sim 10^3 \mathrm{s}^{-1}$, and above their non-radiative relaxation rate at $10$\,mK. These conditions can likely be met in a variety of paramagnetic systems, ranging from organic radicals to transition-metal ions in metallo-enzymes, which could be deposited on top of the resonator constriction in the form of small molecular crystals or even frozen solutions. Applying the techniques demonstrated here may then yield their hyperfine couplings to individual ligand nuclear spins with a higher spectral resolution than reached with ensemble measurements.

\subsection*{Authors contribution}
{The experiment was designed by J.T., J.O'S., E.F., and P.B. The crystal was grown by P.G. and characterized by EPR spectroscopy by S.B. The spin resonator chip was designed and fabricated by J.T. with the help of P.A. The SMPD was designed, fabricated, and characterized by L.P. under supervision of E.F. Data were acquired by J.T., J.O'S., with the help of Z.W.H., and P.H. Data analysis and simulations were conducted by J.T., J.O'S., Z.W.H., and P.H. The manuscript was written by J.T. and P.B., with contributions from all co-authors. The project was supervised by E.F. and P.B}

\subsection*{Acknowledgements}
{We acknowledge technical support from P.~S\'enat, D. Duet, P.-F.~Orfila and S.~Delprat, and are grateful for fruitful discussions within the Quantronics group. We acknowledge support from the Agence Nationale de la Recherche (ANR) through the MIRESPIN (ANR-19-CE47-0011) project. We acknowledge support of the R\'egion Ile-de-France through the DIM QUANTIP, from the AIDAS virtual joint laboratory, and from the France 2030 plan under the ANR-22-PETQ-0003 grant. This project has received funding from the European Union Horizon 2020 research and innovation program under the project OpenSuperQ100+, and from the European Research Council under the grant no. 101042315 (INGENIOUS). We thank the support of the CNRS research infrastructure INFRANALYTICS (FR 2054) and Initiative d'Excellence d’Aix-Marseille Université – A*MIDEX (AMX-22-RE-AB-199). We acknowledge IARPA and Lincoln Labs for providing the Josephson Traveling-Wave Parametric Amplifier. We acknowledge the crystal lattice visualization tool VESTA.}

\appendix

\section{Sample and experimental setup}
\label{section:sample}

The CaWO$_4$ sample was cut out of a boule grown at Institut de Recherche de Chimie Paris in a rectangular slab shape, same as the one used in \cite{wang_single-electron_2023}. It's dimensions are 7x4x0.5 mm, with its surface approximately along the $(a,c)$ plane of the crystal and the $c$ axis along the short edge. The gyromagnetic tensor of Er$^{3+}$ is diagonal in the $(a, b, c)$ coordinate system, $\gamma_a=\gamma_b=\gamma_\perp= - 2\pi\cdot117.3
$ GHz/T and $\gamma_c=\gamma_\parallel= -  2\pi\cdot17.45$ GHz/T \cite{antipin_a._paramagnetic_1968}.

\begin{figure}[b]
    \centering
    \includegraphics[width=\columnwidth]{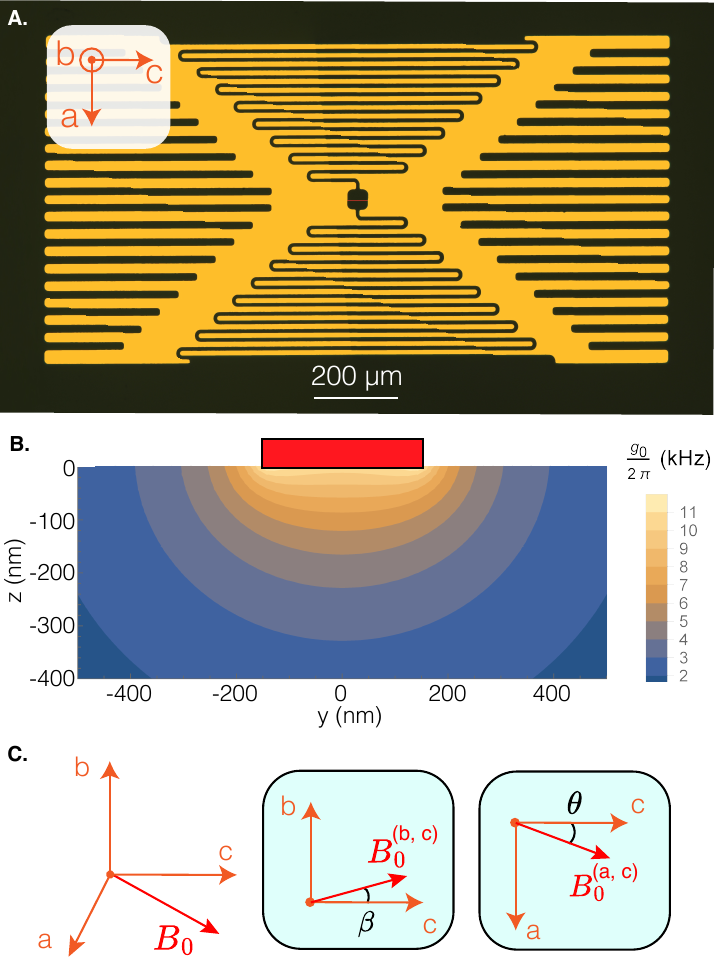}
    \caption{
        \textbf{Micrograph of the sample and magnetic field orientation structure}. 
        \textbf{A}. False color micrograph of a Nb thin-film resonator of the same design fabricated on a CaWO$_4$ crystal. The nanowire is colored in red while the rest of the resonator is colored yellow. The picture is stitched from 2 independent images.
        \textbf{B}. Spatial map for the coupling $g_0$ between the Er$^{3+}$ and the superconducting resonator. The map is shown as a transversal cut around the nanowire (in yellow).
        \textbf{C}. Crystalline axis and magnetic field. $\beta$ and $\theta$ are defined as the angle between the $c$ axis and the projection of $B_0$ in the $(b, c)$ plane and the $(a, c)$ plane respectively.   
    }
    \label{fig:sample}
\end{figure}

The superconducting resonator was fabricated on top of the crystal. The process starts with the sputtering of a 50 nm thin-film of niobium on top of the CaWO$_4$ sample. A 30 nm aluminum mask, in the resonator shape, is deposited on top of the niobium with a lift-off process. The unmasked niobium is removed through dry etching using a 1:2 mix of CF$_4$ and Ar gas. Finally, the Al mask is removed in a MF-319 solution. Figure \ref{fig:sample}A shows a stitched, false color micrograph of a resonator.

\begin{figure*}[t]
    \centering
    \includegraphics[width=\textwidth]{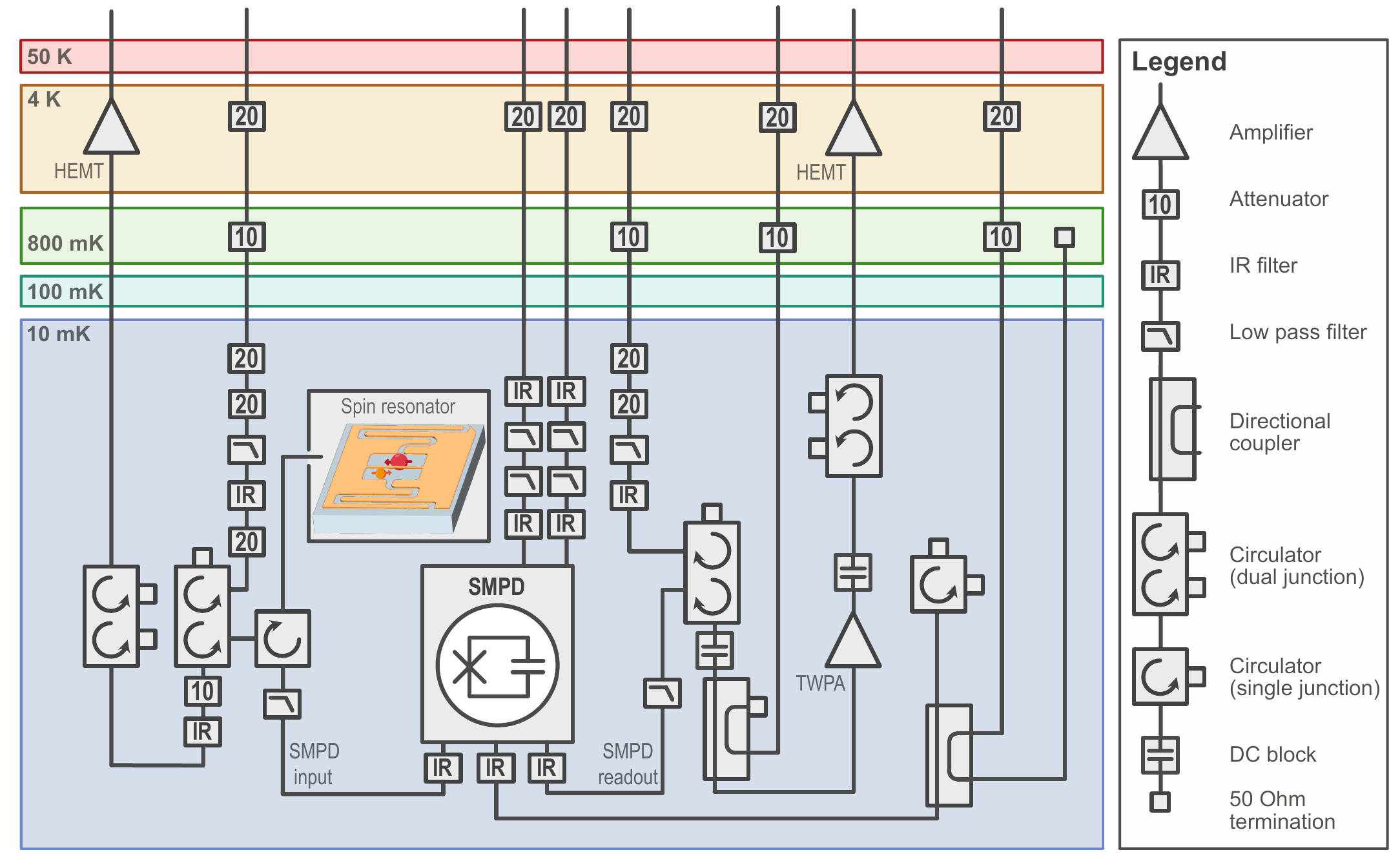}
    \caption{
        \textbf{Schematic of the setup}. Wiring and components of the experimental setup within the dilution fridge.
    }
    \label{fig:wiring}
\end{figure*}
Three resonators with a frequency separation of 150 MHz are fabricated on the sample. Resonator frequencies were measured, and reworked by etching part of the capacitance to lie in the SMPD operating range (7.70 - 7.76~GHz). During the rework process of the sample used for this experiment, the crystal broke and two resonators were destroyed. Without applied field, the remaining resonator's frequency is 7.774 GHz, $Q_{\mathrm{ext}}$= $1.6 \cdot 10^4$ and $Q_{\mathrm{int}}$ = $4.0 \cdot 10^4$. With $B_0$ = 446 mT approximately parallel to the crystalline $c$-axis, the resonator's frequency is 7.748 GHz, $Q_{\mathrm{ext}}$= $2.0 \cdot 10^4$ and $Q_{\mathrm{int}}$ = $2.8 \cdot 10^4$.

The sample is hosted in a 3D copper cavity with a single port which couples capacitevely to the resonator through a microwave antenna The full wiring diagram is shown in Fig. \ref{fig:wiring}. The resonator uses a the same "bowtie" design as described in \cite{wang_single-electron_2023}, albeit with a reduced 0.3 x 50~$\upmu$m wire. The reduced dimensions of the wire result in an increased coupling between the Er$^{3+}$ ions and the resonator, represented as a 2D map in \ref{fig:sample}B. The sample holder is thermalized at $10$\,mK in a dilution cryostat, and placed at the center of a vector 1/1/1 T magnet. The magnetic field $B_0$ must be applied in the resonator $(a,c)$ crystalline plane, since out-of-plane components will rapidly reduce the quality factor and destroy the superconducting properties of the resonator. We align the magnetic field in the resonator plane, by using the resonator response as described in \cite{wang_single-electron_2023}. However, the resonator plane is not exactly aligned with the the $(a,c)$ plane. We define $\beta$ the angle between the resonator plane and the $c$-axis, and $\theta$ the angle between the $c$-axis projection and the applied magnetic field (see Fig. \ref{fig:sample}C)

\section{SMPD characteristics}
\label{section:smpd}

All measurements were performed with the use of a Single Microwave Photon Detector (SMPD). Succinctly, the arrival of a microwave photon at the device input is mapped onto the excited state of a superconducting transmon qubit by a parametric process activated by a pump tone. The qubit is dispersively read-out cyclically and counts are recorded when found in the excited state. More details about the design and previous versions of the device can be found in \cite{albertinale_detecting_2021, balembois_cyclically_2024}. The counter used in the experiment was operated with a bandwidth of $\sim 300$ kHz for which it presents a maximum efficiency of 0.79(1) and a dark count rate of 40(5) counts per second. The average duration of a complete measurement cycle is $\sim 17 \mu$s, with a certain variation due to the active reset performed on the qubit, which can extend the sequence. The down-time of the detector, during which the qubit is read-out, is $\sim 2~\upmu$s per cycle. 

\begin{figure}[t]
    \centering
    \includegraphics[width=\columnwidth]{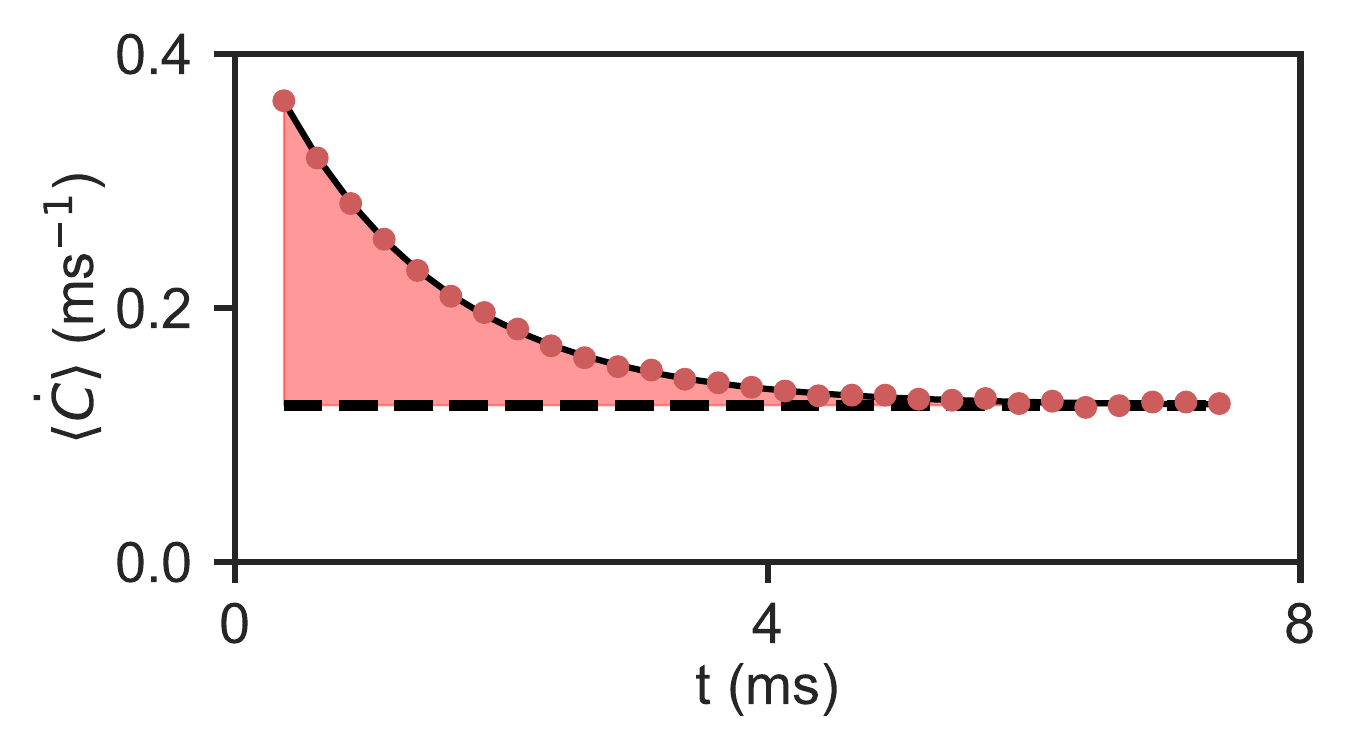}
    \caption{
        \textbf{Spin-efficiency and dark-counts measurement for $\mathrm{Er}_5$}.
        Measured (dots) fluorescence count rate is fitted with an exponential decay (solid black line). The offset of the decay corresponds to the SMPD dark-count rate $\Gamma_{DC}=0.12$ counts/ms (dashed black line). The spin-efficiency $\eta=0.4$ is the integral (red area) of the exponential decay .
    }
    \label{fig:efficiency}
\end{figure}

The operational dark-counts, $\Gamma_{DC}$, and the spin-efficiency $\epsilon$, probability of measuring a photon coming from the spin sample, were measured through the fluorescence decay of a single spin (see Fig. \ref{fig:efficiency}). The average number of photons detected from a spin signal is measured by integrating the background-removed fluorescence curve. Since the number of spins being excited is one, the average number of expected photons is equivalent to the spin-efficiency. Furthermore, the signal will present a flat background count-rate corresponding to the dark counts of the detector, coming from spurious excitations in the qubit and heat dissipation on the microwave lines. During the measurements, the spin-efficiency ranged from 0.2 to 0.4 and the dark count rate spanned from 40 to 150 counts per second, depending on the quality of the calibration of the detector. The spin efficiency does not reach the maximum efficiency of the SMPD due to the internal losses of the resonator.

\section{Tracking algorithms}
\label{section:tracking}


As is visible in Fig.2 of the main text, the allowed EPR frequencies of individual erbium ions shows fluctuations over time, due to a combination of $B_0$ drifts, nuclear spin environment dynamics, and charge noise. The measured drift is comparable to the linewidth of the EPR-resonance of the Er$^{3+}$ ions. However the drift is slower than cross-relaxation events and can be compensated for. Two different tracking algorithms were implemented, one based on direct spectroscopy and the other a continuous Proportial-Integral (PI) loop. 

\begin{figure}[t]
    \centering
    \includegraphics[width=\columnwidth]{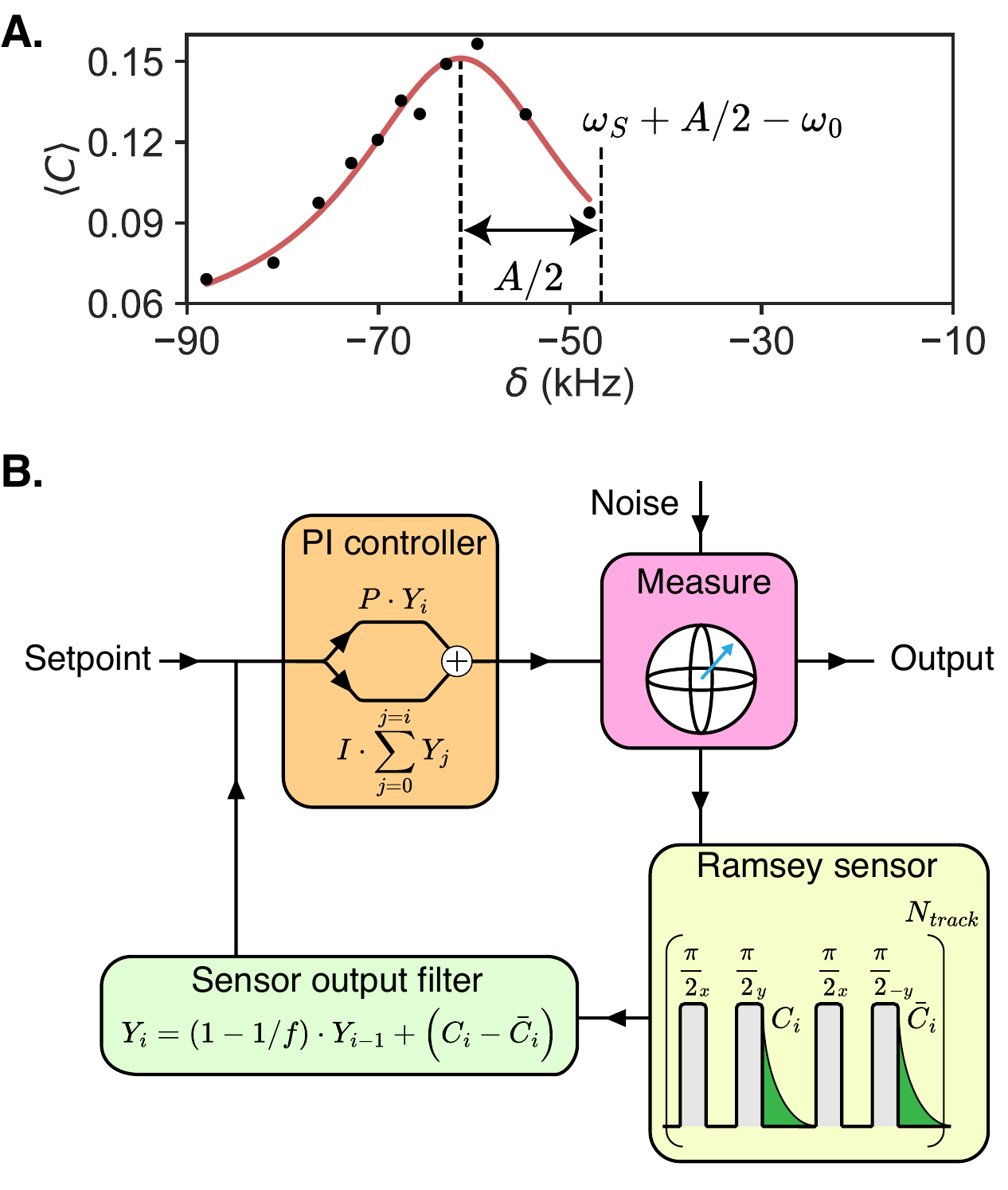}
    \caption{
        \textbf{Spectral tracking measurement and feedback loop diagram}.
        \textbf{A}. Spectral tracking sample measurement. Measured (dots) ensemble averaged number of counts as a function of $\delta$ after pumping the zero-quantum transition of $\mathrm{Er}_5$. A Lorentzian fit (solid line) yields the center of the EPR resonance (left dashed vertical line). $\omega_S$ is calculated by subtracting $A/2$ (right dashed vertical line). 
        \textbf{B}. Feedback loop diagram. The PI loop is run after every measurement to correct any shift in the frequencies of the system. After a given iteration of the measurement $i$, the Ramsey sensor measures $C_i$ - $\bar{C}_i$, a proxy quantity for the detuning of the system. The sensor is filtered to obtain $Y_i$. Finally, the detuning of the system is corrected via a PI controller before the next measurement starts.
    }
    \label{fig:tracking}
\end{figure}

The first algorithm performs high-resolution spectroscopy with a 80 $\mu$s $\pi$-pulse in a 50 kHz range centered around the expected transition frequency. If there are multiple possible transitions due to nearby nuclear spins, the nuclear spins are initialized into a predetermined state before the spectroscopy. The ensemble averaged number of counts $C(\delta)$ is then fitted to a Lorentzian model, as shown in Figure \ref{fig:tracking}A. From the center of the fit, the drift is calculated. The measurement takes 1 minute to complete. Due to its long duration, it can only be run a limited number of times, which results in a poor tracking.

The second method uses a PI-loop to track the value of one of the electron spin transitions. A diagram explaining the control sequence is shown in \ref{fig:tracking}B. After performing a measurement, noise from the environment induce random drifts. This drift can be characterized through a Ramsey experiment. If the drive frequency is detuned $\Delta \omega$ from the transition frequency, the state of the system after a Ramsey sequence is $\langle S_z \rangle = e^{-t/T_2^x}\cdot \sin{(\Delta\omega\cdot \tau)}$. The number of fluorescent counts $C$ gives a proxy for $\langle S_z \rangle$, making this sequence a good sensor for $\Delta \omega$. However, the presence of fluctuating dark counts in the measurement makes defining a setpoint for the PI controller a hard task. As such, we effectively remove this problem by performing the negative phase Ramsey sequence and subtracting the measured number of counts $\bar{C}$ from the original value. The two measurements are interleaved and performed $N_{track}=10$ times. The quantity $C - \bar{C}$ is a good sensor since it has a stable setpoint at 0 and depends linearly with the detuning, as long as $\tau \ll 2\pi/\Delta\omega$. To increase the convergence speed, we allow the system to have a short memory time by filtering the values of the sensor output for every iteration $i$ of the PI loop to generate the filtered sensor $Y_i$ 

\begin{equation}
\begin{split}
    Y_i &= \sum_{j=0}^i \left( C_j - \Bar{C}_j \right)\cdot e^{\sfrac{-(j-i)}{f}} \approx \\ 
    &\approx (1-\sfrac{1}{f}) \cdot Y_{i-1} + \left( C_i - \Bar{C}_i \right)
\end{split}
\end{equation}

Where $f=2000$ is the memory of the system. Finally, the PI controller calculates the frequency correction to be applied on the pulses through a proportional and an integral channel

\begin{equation}
    \delta \omega_i = P \cdot Y_i + I \cdot \sum_{j=0}^i Y_j.
\end{equation}

The values for $P$ and $I$ were set by manually optimizing the convergence time after artificially detuning the pulse frequency. As for the other sequence, we initialized any resolved nuclear spins into an predefined state before the tracking is performed.

\section{Time trace analysis}
\label{section:traces}

\begin{figure*}[t]
    \centering
    \includegraphics[width=0.7\textwidth]{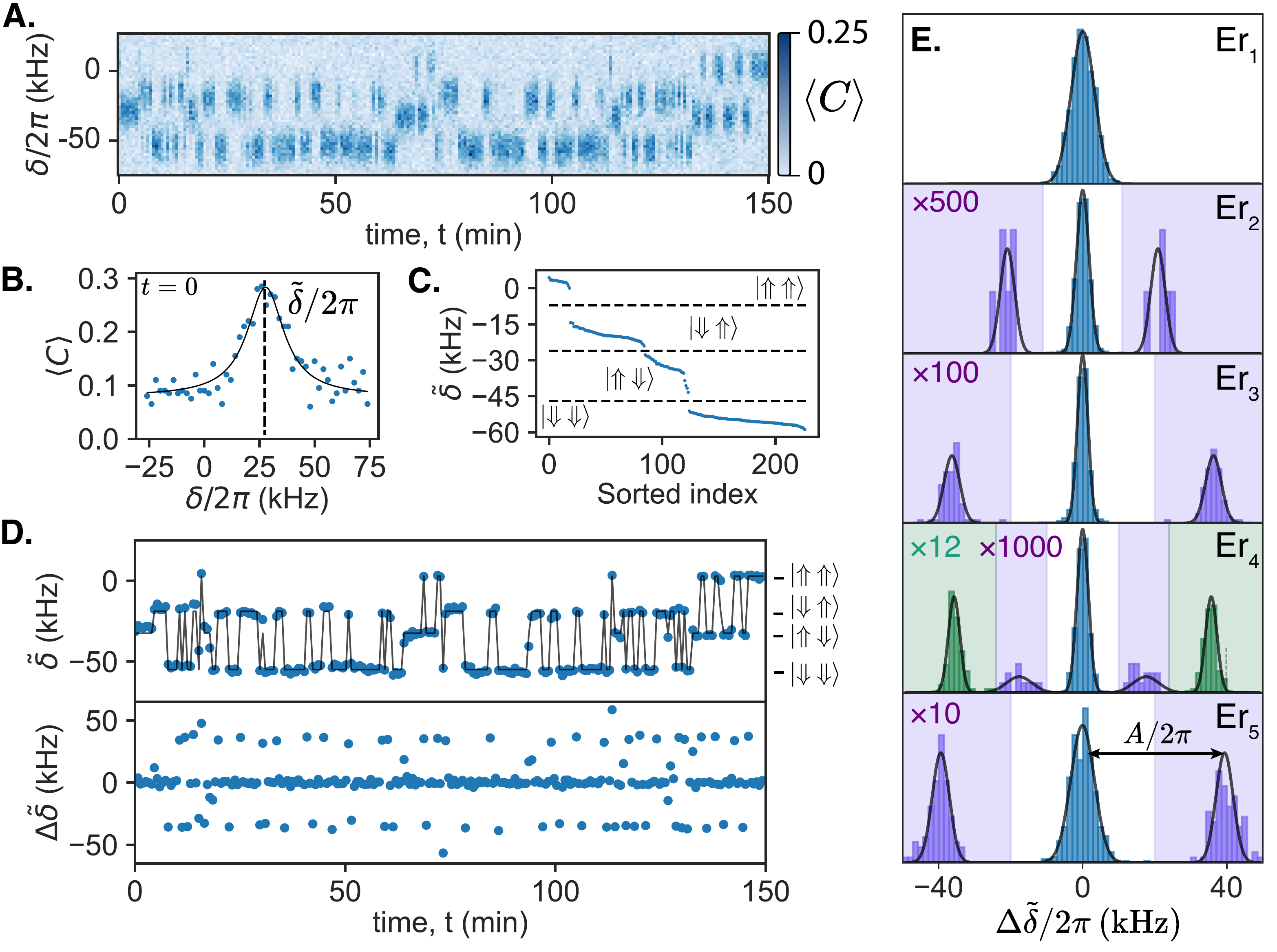}
    \caption{
        \textbf{Trace analysis steps}. 
        \textbf{A.} High resolution spectra as a function of time for $\mathrm{Er}_4$.
        \textbf{B.} Spectrum at time equal to 0 s (blue dots) and Lorentzian fit (black line) with center $\Tilde{\delta}$ (dashed black line).
        \textbf{C.} Sorted values of $\Tilde{\delta}$ (blue dots) and derivative bound thresholds (dashed black lines). Each sector corresponds to the nuclear spins state that is indicated.
        \textbf{D.} \textit{top}. Lorentzian fit center $\Tilde{\delta}$ (blue dots) as a function of time and nuclear spin state assignment (black line).\textit{bottom}. Difference between consecutive values of $\Tilde{\delta}$ as a function of time.
        \textbf{E.} Histograms of $\Delta\tilde{\delta}$ for the five ions. The colored areas have been enhanced by a factor as indicated. Solid lines are Gaussian fits to each peak. The difference between their centers yields the isotropic hyperfine coupling constant of each nuclear spin, $A/2\pi$.
    }
    \label{fig:traces}
\end{figure*}

The traces by performing high-resolution spectroscopy with an 80 $\upmu$s long Gaussian pulse and sweeping the frequency over 100~kHz, then the ensemble average number of counts is recorded. The duration of the detection window after the pulse is set to approximately the $T_1$ of the spin. After the pulse is sent, we wait between 60 and 120 $\mu$s to avoid spurious counts due to the heating of the microwave lines. Each curve from the trace is calculated by averaging 200 frequency sweeps. A trace for $\mathrm{Er}_5$ is presented in Fig \ref{fig:traces}A. Each each curve is fitted with a multi-Lorentzian model and the center of the most prominent resonance is defined as $\Tilde{\delta}$ (Fig. \ref{fig:traces}B). To determine the state of the nuclear spin state for each moment in time, all the values of $\Tilde{\delta}$ are sorted in increasing order and thresholds are calculated based on a derivative bound (see Fig. \ref{fig:traces}C). The state of the nuclear spin is then classified based on the frequency of the resonance (see Fig \ref{fig:traces}D). We also define $\Delta\tilde{\delta}$, as the difference between consecutive values of $\tilde{\delta}$, $\Delta\tilde{\delta}(t)\equiv\tilde{\delta}(t) - \tilde{\delta}(t-\Delta t)$.

Histograms of $\Delta \tilde{\delta}$ for $\mathrm{Er}_{1-5}$ are shown in Fig.~\ref{fig:traces}E, which consist of several peaks. The peak centered at $0$ arises when no resolved jump occurs in-between two consecutive traces. A Gaussian fit yields a Full-Width-Half-Maximum varying from 4.5 to 10 kHz, indicating that the short-term noise (at the minute scale) also varies significantly from ion to ion. The extra peaks observed for $\mathrm{Er}_{2-5}$ arise from resolved nuclear spin quantum jumps. Gaussian fits to these peaks yield the mean frequency jump size which we identify to the isotropic hyperfine coupling $A$, and the jump standard deviation which we find similar to the central peak, as expected. 

Using the state assignment, the cross-relaxation probabilities $\eta^{d, z}$, is measured. $\eta^{d, z}$ is defined as

\begin{equation}
    \eta^{d, z} \approx \frac{\Gamma_x^{d,z}}{\Gamma_R } \approx \frac{N_x^{d, z}}{N_{\mathrm{exc}}^{\Downarrow, \Uparrow}},
\end{equation}

where $N_x^{d, z}$ is the number relaxation events through the zero- and double-quantum transition and can be directly counted from the now state classified traces. $N_{\mathrm{exc}}^{\Downarrow, \Uparrow}$ is the number of excitations of the electron spin when the nuclear spin is in the $\Downarrow$ (resp. $\Uparrow$) state. This quantity is directly proportional to the number of curves classified in the respective nuclear spin state, while taking into consideration the relative amplitudes of the multi-Lorentzian fit. This value is then multiplied by the number of averages of each spectrum. However, due to the non-negligible spectral width of the Gaussian pulse, each sweep does not excite the electron spin exactly one time. Based on the shape and length of the pulse as well as the frequency step of each measurement, an approximate number of excitation pulses is calculated from the Fourier transform of the pulse. Table I presents the cross-relaxation probabilities for the coupled nuclear spins of $\mathrm{Er}_{2-5}$.

\begin{table}[h]
\begin{tabular}{|c|c|c|c|c|c|}
\hline
                        & \textbf{$\mathrm{Er}_2$}    & \textbf{$\mathrm{Er}_3$}   & \textbf{$\mathrm{Er}_4^{(1)}$} & \textbf{$\mathrm{Er}_4^{(2)}$}  & \textbf{$\mathrm{Er}_5$}  \\ \hline
$\eta^d \cdot 10^{5}$   & 1.5$\pm$7             & 5.8$\pm$6            & 18$\pm$2                  & 1.6$\pm$7                & 24$\pm$2            \\ \hline
$\eta^z \cdot 10^{5}$   & 0.8$\pm$3             & 4.2$\pm$4            & 9$\pm$1                   & 0.7$\pm$3                & 24$\pm$2            \\ \hline
\end{tabular}
\label{tab:cross_relaxation}
\caption{Cross-relaxation probability measured through the time-trace analysis. The uncertainties of the cross-relaxation probabilities are estimated through the Wald method with a 1 sigma confidence interval.}
\end{table}

In the case of $\mathrm{Er}_{2}$ and $\mathrm{Er}_{4}$ we measure a significant difference between $\eta^{d}$ and $\eta^{z}$, the former being almost double than the latter. This is consistent with the behaviour that the nuclear spins display in the time trace, spending a considerably longer amount of time in the $\Downarrow$ state. To understand the origin of this difference, more in-detail measurements of the system are required. A possible explanation is the existence of an extra, non-radiative relaxation channel.

\section{Read-Out probability fit}
\label{section:readout}

The probability of correctly assigning the state of the nuclear spin after $N_{RO}$ measurements is obtained from a combination of two effects: the increase of the signal-to-noise ratio (SNR) as a function of the number of measurements and the evolution of the nuclear spin states populations due to cross-transition events.

Let us consider the measurement of a single nuclear spin that is completely polarized in $|\Downarrow\rangle$. As detailed in the main text, $C_\Uparrow$ and $C_\Downarrow$ are the total number of counts measured after applying a $\pi$ pulse and an SMPD measurement $N_{RO}$ times at the two EPR-allowed frequencies of the electron spin. Since the nuclear spin is completely polarized, the only contributions to $C_\Uparrow$ will be the dark counts of the detector, which results in a Poissonian distribution with rate $\lambda = N_{RO} \cdot \Gamma_{DC}\cdot t_D$ where $\Gamma_{DC}$ is the dark count rate of the detector and $t_D$ is the duration of the detection window. On the other hand, $C_\Downarrow$ will be the result of two processes, the aforementioned Poissonian events due to dark counts and the real detection events, which follow a binomial distribution with $n = N_{RO}$ and  probability $p = \epsilon$, where $\epsilon$ is the probability of detection. The direct addition of the two distributions is not straightforward, however, both the Poisson and binomial distributions can be approximated to a Gaussian distribution when $\lambda >> 1$ and $n>>1$ respectively. Under this approximation the addition of the two Gaussian distributions is trivially performed by adding the means and variances. The distributions for $C_\Uparrow$ and $C_\Downarrow$ are then

\begin{equation}
\begin{split}
    &C_\Uparrow \sim \mathcal{N} (\mu_{DC}, \sigma^2_{DC}) \\
    &\mu_{DC} = \sigma^2_{DC} = N_{RO} \cdot \Gamma_{DC}/2\pi\cdot t_D
\end{split} 
\end{equation}

\begin{equation}
\begin{split}
    &C_\Downarrow \sim \mathcal{N} (\mu, \sigma^2) \\
    &\mu = N_{RO} \left(\epsilon + \Gamma_{DC}\cdot t_D \right) \\
    &\sigma^2 = N_{RO} \left(\epsilon \cdot (1 - \epsilon) +  \Gamma_{DC}/2\pi\cdot t_D \right)
\end{split}
\end{equation}

The difference in counts $\delta C = C_\Downarrow - C_\Uparrow$ is used to measure the state of the nuclear spin. The integral between 0 and $\infty$ of the distribution for $\delta C$ is a direct measure of $P\Downarrow$. Since $C_\Downarrow$ and $ C_\Uparrow$ are Gaussian distributions,

\begin{equation}
\begin{split}
    \delta C &\sim \mathcal{N} (\mu - \mu_{DC}, \sigma^2 + \sigma^2_{DC}) \\     
    P_\Downarrow &= \int_0^\infty \frac{e^{-\frac{(x - (\mu - \mu_{DC}))^2}{2\cdot\left(\sigma^2 + \sigma^2_{DC}\right)}}}{\sqrt{2\pi\cdot\left(\sigma^2 + \sigma^2_{DC}\right)}} dx = \\
    &= \frac{1}{2}\left(1 - \text{Erf}\left[\frac{\text{SNR}}{\sqrt{2}}\right]\right) \\
    \text{SNR} =& \frac{\epsilon}{\sqrt{\epsilon \cdot (1 - \epsilon) +  2\Gamma_{DC}/2\pi\cdot t_D}} \cdot \sqrt{N_{RO}}
\end{split}
\end{equation}

The SNR increases with $\sqrt{N_{RO}}$ which leads to an exponential increase of the probability of a correct detection with $N_{RO}$. The prefactor is calculated from the fluorescence curve of the same measurement, from which the following parameters were obtained, $\epsilon = 0.18$, $\Gamma_{DC} = 150$ s$^{-1}$, $t_D = 2.6$ ms. More details about how these values are obtained are given in section \ref{section:smpd}.

The cross-relaxation events introduce a dynamic change in the population of the nuclear spin states. Due to the finite probability $\eta^{z,d}$ to relax into the opposite state, the evolution of the population can be modeled via the following rate equations

\begin{equation}
\begin{split}
    \frac{\Delta p_{|\Downarrow\rangle}}{\Delta N_{RO}} \: = \: &\eta^d p_{|\Uparrow\rangle} - \eta^z p_{|\Downarrow\rangle}, \\
    \frac{\Delta p_{|\Uparrow\rangle}}{\Delta N_{RO}}  \: = \: &\eta^z p_{|\Downarrow\rangle} - \eta^d p_{|\Uparrow\rangle}, \\
    p_{|\Uparrow\rangle} + &p_{|\Downarrow\rangle} = 1.
\end{split}
\end{equation}

In order to account for preparation errors the initial condition $p_{|\Downarrow\rangle}(0) = p_0 < 1$ is used. Since $\eta^{z,d} \ll 1$, the dynamics of the system are much slower than the step size of $N_{RO}$ and it can be approximated to $\frac{\Delta p_{|\Downarrow\rangle}}{\Delta N_{RO}} \approx \frac{d p_{|\Downarrow\rangle}}{d N_{RO}}$. Solving the equations 

\begin{equation}
    p_{|\Downarrow\rangle}(N_{RO}) \: = \: (p_0 - \frac{\eta^d}{\eta^d + \eta^z})\cdot e^{-(\eta^d + \eta^z)\cdot N_{RO}} + \frac{\eta^d}{\eta^d + \eta^z}, 
\end{equation}

In the case of $\mathrm{Er}_5$, $\eta^z \approx \eta^d = \eta$ (see App.\ref{section:traces}) and 

\begin{equation}
    p_{|\Downarrow\rangle}(N_{RO}) \: \approx \: (p_0 - \sfrac{1}{2})\cdot e^{- 2\eta\cdot N_{RO}} + \sfrac{1}{2}, 
\end{equation}

Since SNR$(N_{RO}) \gg 2\eta\cdot N_{RO}$ the two processes can be considered independent and the total fidelity of the read-out can be approximated to

\begin{equation}
    P_\Downarrow \approx \frac{1}{2}\left(1 - \text{Erf}\left[\frac{\text{SNR}}{\sqrt{2}}\right]\right) \cdot p_{|\Downarrow\rangle}(N_{RO})
\end{equation}

An equivalent expression can be found for the complementary process when preparing into $|\Uparrow\rangle$. Fitting the model to the experimental data we obtain $p_0=0.97\pm0.02$ and $\eta=3.2\pm0.2 \cdot 10^{-4}$ for the cross-relaxation $|\Downarrow\rangle \rightarrow |\Uparrow\rangle$. For the complementary transition, $p_0=0.94\pm0.02$ and $\eta=2.6\pm0.2 \cdot 10^{-4}$. Using the relation presented in the main text $\eta =  \frac{B^2}{4\omega_I^2 } \frac{\kappa^2}{\kappa^2 + 4\omega_I^2}$ we obtain $B/2\pi=74\pm7$ kHz  and $B/2\pi=67\pm6$ kHz respectively. Within error, the two values are the similar. However, the measurement does not capture the tail of the exponential reducing the accuracy of the result.

\section{Measurement of $\omega_I$}
\label{section:omega_I}

\begin{figure}[t!]
    \centering
    \includegraphics[width=0.95\columnwidth]{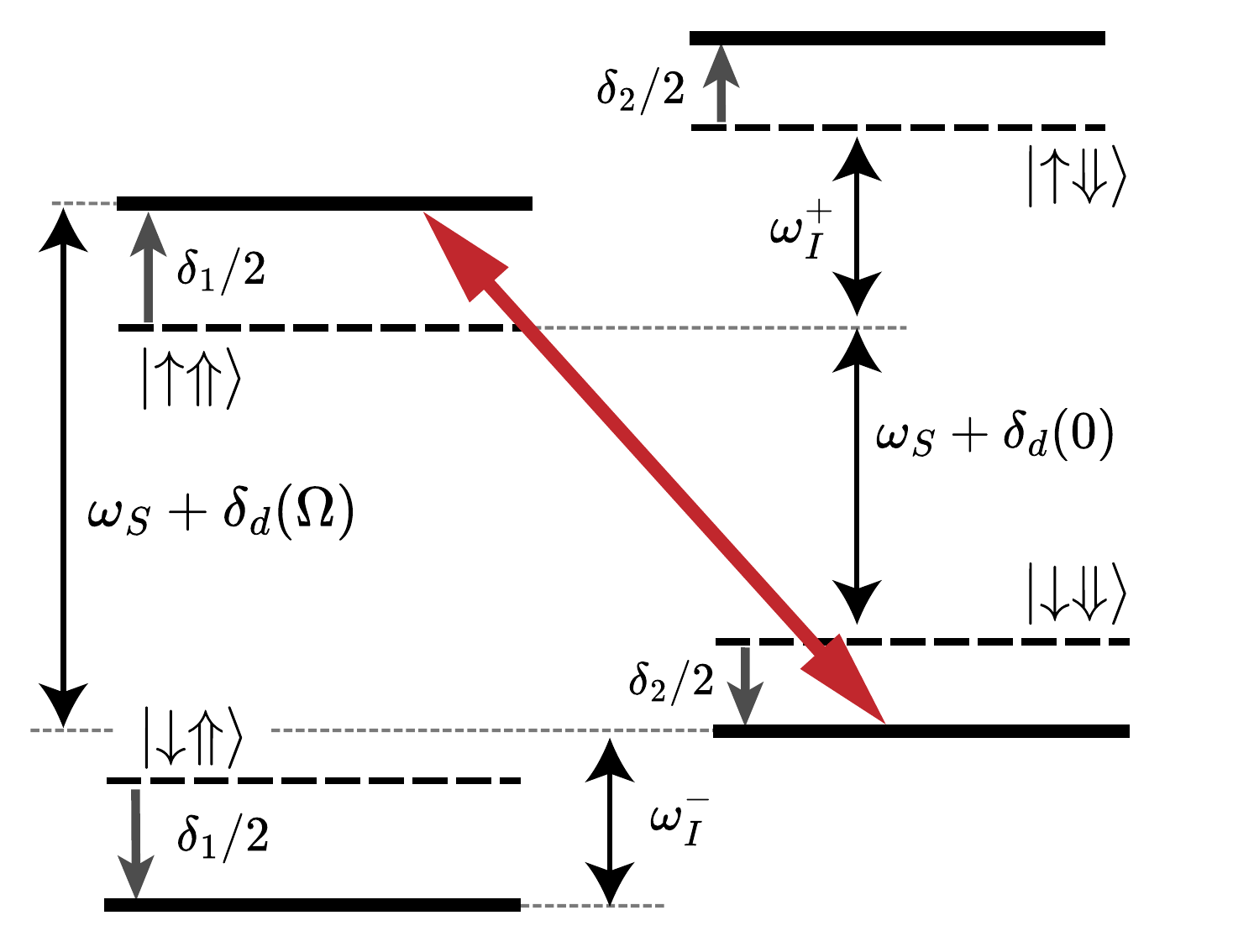}
    \caption{
        \textbf{AC-Zeeman shift of the energy level diagram under off-resonant drive} The spin levels are frequency-shifted by the microwave drive, from their $\Omega=0$ value (black dashed lines) to the value under drive (solid black line). Microwave drive of amplitude $\Omega_d$ is shown as a red double arrow.
    }
    \label{fig:zeeman}
\end{figure}

\begin{figure*}[t!]
    \includegraphics[width=0.9\textwidth]{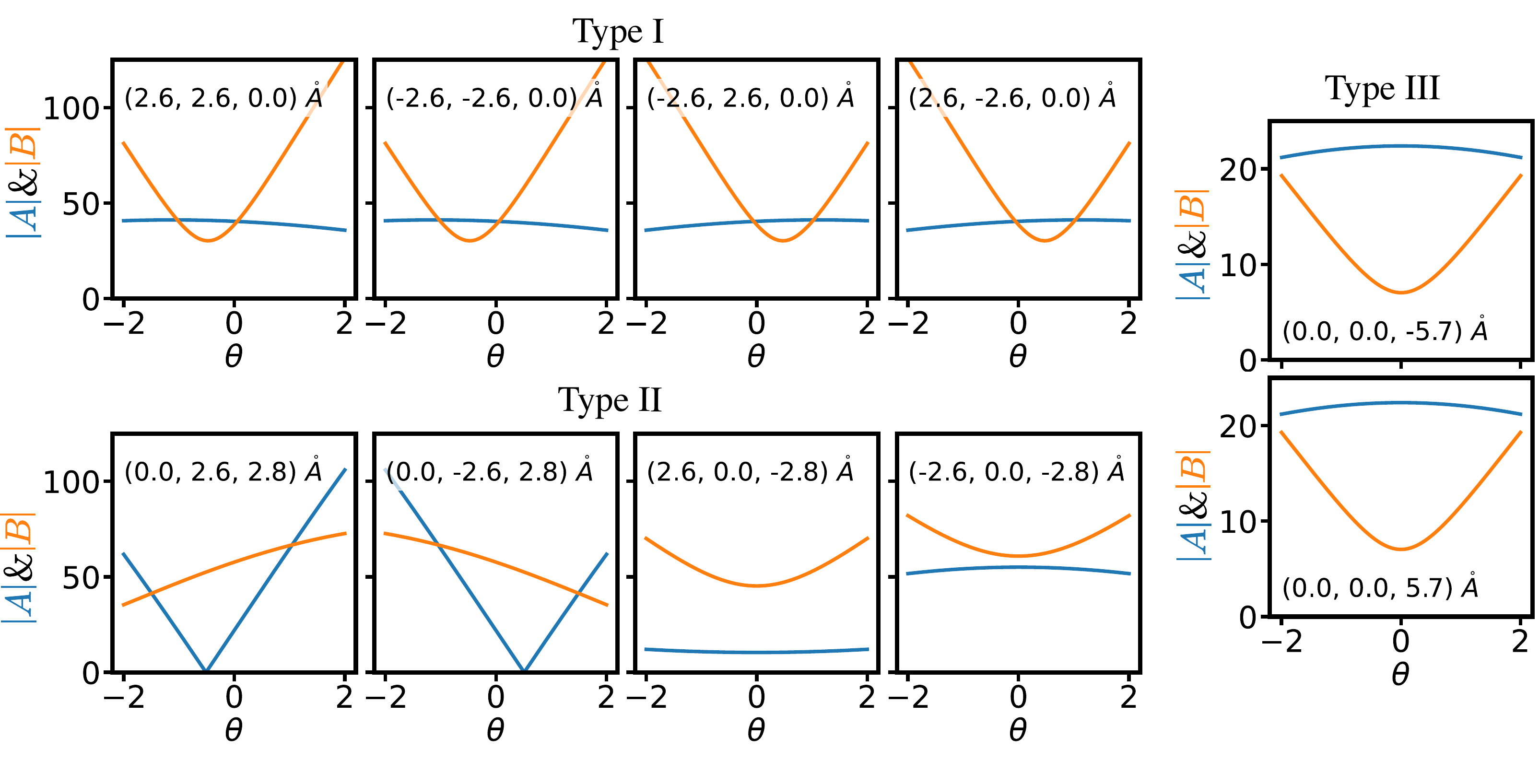}
    \caption{
        \textbf{Hyperfine parameters for Types I, II and III nuclear spins as a function of $\theta$.}
        Hyperfine coupling $A$ and $B$ obtained through a dipole-dipole Hamiltonian simulation. The values are shown as a function of the angle between $B_0^{(a,c)}$ and the $c$ axis of the crystal, $\theta$. The simulation considers the measured misalignment angle between $B_0^{(b,c)}$ and the $c$ axis of the crystal $\beta=0.8$. The position of the ions is specified in angstroms. The plots are organized depending on the relative position of the nuclear spin with respect to the Er$^{3+}$ ion.
    }
    \label{fig:hyperfine}
\end{figure*}

To measure the bare nuclear spin frequency $\omega_I$, we rely on the observed AC-Zeeman shift when driving the forbidden transitions. The undriven frequency difference $\delta_{d, z}^{(0)}$ between $\omega_S$ and double and zero-quantum transition frequency can be analytically obtained from the Hamiltonian,

\begin{equation}
\begin{split}
     \delta_{d,z}^{(0)} =\pm \frac{1}{2}&(\omega^+_{I} + \omega^-_{I}), \\
     \omega^+_{I} = (\omega_I+\sfrac{A}{2})&\cos\xi_+ - \sfrac{B}{2}\sin\xi_+, \\
     \omega^-_{I} = (\omega_I-\sfrac{A}{2})&\cos\xi_-  + \sfrac{B}{2}\sin\xi_-, \\
     \tan \xi_+ = &\frac{-B}{A+2\omega_I}, \\
     \tan \xi_-  = &\frac{-B}{A-2\omega_I}.
\end{split}
\end{equation}

Which in the high-field limit $\omega_S\gg A,B$, this approximates to $\delta_{d,z}^{(0)}\approx\omega_I$. When resonantly driving the forbidden transitions, the energy levels will be shifted due to the AC-Zeeman effect. The origin of the shift lies in the frequency difference between the drive and the allowed transitions and depends on the amplitude of the drive $\Omega$.

In the case of a simple two-level system, the ac-Zeeman shift is given by $\sfrac{\Omega^2}{2\Delta}$ where $\Delta$ is the detunning of the drive with respect to the transition. 
In the case of a four-level system, there are two allowed transitions, each detuned by $\Delta_{1, 2}$. Each of the two allowed transitions is in turn  shifted by $\delta_{1,2} = \sfrac{\Omega^2}{2\Delta_{1,2}}$. The frequency of the forbidden resonance under drive is therefore

\begin{equation}
\begin{split}
     \delta_{z,d}(\Omega) & = \delta_{z,d}^{(0)} \pm \frac{1}{2}\left(\delta_1 + \delta_2\right)\\
     & = \delta_{z,d}^{(0)} \pm \frac{1}{2}\left(\frac{\Omega^2}{2\Delta_{1}} + \frac{\Omega^2}{2\Delta_{2}}\right),
\end{split}
\end{equation}

For a resonant drive on the forbidden transition $\Delta_{1, 2} = \omega^\pm_I - \sfrac{(\delta_1 + \delta_2)}{2}$ (see Fig. \ref{fig:zeeman}). Rearranging the terms we obtain that the center of the resonance is

\begin{equation}
\begin{split}
    \delta_{z,d}(\Omega) =\delta_{z,d}^{(0)} + &\frac{\Omega^2}{4} \frac{1}{\omega^+_{I} - \delta_{z,d}(\Omega) + \delta_{z,d}^{(0)}} \\
    + &\frac{\Omega^2}{4} \frac{1}{\omega^-_{I} - \delta_{z,d}(\Omega) + \delta_{z,d}^{(0)}}.
\end{split}
\end{equation}

Evaluating the expression we obtain

\begin{equation}
\begin{split}
    \delta_{z,d} = \pm \delta_{z,d}^{(0)} - &\frac{\Omega^2}{2} \left( \frac{1}{2\delta_{z,d} + \Delta\omega_I} + \frac{1}{2\delta_{z,d} - \Delta\omega_I} \right), \\
    \Delta&\omega_I = \omega^+_{I} - \omega^-_{I}.
\end{split}
\end{equation}

The experimental data presented in Fig. \ref{fig4}E was fitted to this model, using the measured values for $\mathrm{Er}_5$ of $A/2\pi=34.5$ kHz and $B/2\pi=103$ kHz (see App.\ref{section:rabi}), which yields the nuclear spin frequency $\omega_I/2\pi = 788.1(4)$ kHz.

\section{Measurement of B via Rabi frequencies}
\label{section:rabi}

The Rabi frequency for the allowed and forbidden transitions are simply the product of the matrix element of the driving term. Calling ${B_1}$ the drive amplitude at resonance, the allowed Rabi frequency is given by 

\begin{equation}
    \Omega = |\bar{\bar{\gamma}} \cdot \bar{B}_1| \cdot \langle \downarrow \Uparrow | S_x | \uparrow \Uparrow \rangle = |\bar{\bar{\gamma}} \cdot \bar{B}_1| / 2,
\end{equation}

\noindent since the allowed transitions are resonant with the resonator. On the other hand, the zero- and double-quantum frequencies are detuned by $\delta$. Assuming constant input power, the forbidden Rabi frequency is

\begin{equation}
\begin{split}
    \Omega_{z,d} &= |\bar{\bar{\gamma}} \cdot \bar{B}_1| \cdot \langle \downarrow \Uparrow | S_x | \uparrow \Downarrow \rangle \cdot \frac{1}{\sqrt{1+\sfrac{4\delta^2}{\kappa^2}}} \\
    &= \frac{\Omega}{2} \cdot \left(\frac{B}{2\omega_I - A} + \frac{B}{2\omega_I + A}\right) \cdot \frac{1}{\sqrt{1+\sfrac{4\delta^2}{\kappa^2}}}. \\
    \\
    \\
\end{split}
\label{eq:sideband_rabi}
\end{equation}

Therefore, the value of $B$ can be measured by comparing the Rabi frequencies on the allowed and forbidden transitions at a given input power. For the same drive amplitude, Rabi oscillations on the allowed transition are significantly faster, since the matrix element is larger. In the measurements presented in Fig. 3, the drive amplitude for the forbidden transition was $\alpha^{-1}=6.2$ larger compared to the allowed transition, which needs to be taken into account in the $B$ estimate, as:

\begin{equation}
\begin{split}
    |B| = \alpha\frac{\Omega_{z,d}}{\Omega} \cdot (2\omega_I\pm A) \cdot \sqrt{1+\sfrac{4\delta^2}{\kappa^2}} 
\end{split}
\end{equation}

The Rabi frequency from the zero- and double-quantum transitions measurements (resp. 104$\pm$7~kHz and 102$\pm$7~kHz) agree within error. However, there is a significant difference compared to the values obtained through the cross-relaxation method ($\sim$ 70 kHz). 
This is a more precise method compared to the cross-relaxation analysis presented in App.\ref{section:traces} and \ref{section:readout} as it merely relies on frequency fitting coherent oscillations; therefore, in the calculation performed in App.\ref{section:omega_I}, we take $B/2\pi = 103 \pm7$~kHz, the average between the zero- and double-quantum transitions  calculation.

\section{Dipole-dipole interaction calculation and site assignment}
\label{section:dipole}

\begin{table}
    \centering
    \begin{tabular}{|c||c|c|c||c|}
        \hline
        Type I            & \textbf{$\mathrm{Er}_3$} & \textbf{$\mathrm{Er}_4^{(1)}$}  &\textbf{$\mathrm{Er}_5$} & calc. \\ \hline
        $\theta$ (º)      & -0.2                    & -0.1                          & 0.1                    &  -0.2 -- 0.1\\ \hline
        $|A|/2\pi$ (kHz)  & 36.3$\pm$0.6                     & 35.8$\pm$0.5         & 39.6$\pm$0.5                    &  40\\ \hline
        $|B|/2\pi$ (kHz)  & 27$\pm$3                       & 40$\pm$5               & 103$\pm$3           & 35 -- 48\\  \hline
    \end{tabular}
    
    \vspace*{0.5 cm}
    
    \begin{tabular}{|c||c|c||c|}
        \hline
        Type II            & \textbf{$\mathrm{Er}_3$} & \textbf{$\mathrm{Er}_4^{(1)}$} & calc. \\ \hline
        $\theta$ (º)       & -0.2                    & -0.1                            & -0.3 -- -0.1     \\ \hline
        $|A|/2\pi$ (kHz)   & 36.3$\pm$0.6            & 35.8$\pm$0.5                    & 12 -- 57    \\ \hline
        $|B|/2\pi$ (kHz)   & 27$\pm$3                & 40$\pm$5                        & 45 -- 61     \\ \hline
    \end{tabular}
    
    \vspace*{0.5 cm}
    
    \begin{tabular}{|c||c|c||c|}
        \hline
        Type III           & \textbf{$\mathrm{Er}_2$}  & \textbf{$\mathrm{Er}_4^{(2)}$} & calc. \\ \hline
        $\theta$ (º)       & -0.3                      & -0.1  & -0.3 -- -0.1     \\ \hline
        $|A|/2\pi$ (kHz)   & 21$\pm$1                  & 19$\pm$1     & 23     \\ \hline
        $|B|/2\pi$ (kHz)   & 12$\pm$3                  & 11$\pm$3     & 8     \\ \hline
    \end{tabular}
    
    \label{tab:hyperfine_methods}
    \caption{
        Hyperfine parameters for the different Erbium ions and the angle $\theta$ at which the measurements were taken. Ions are organized as per their Type assignment. Three values are given for $B$ of $\mathrm{Er}_5$, obtained from the spectroscopic trace analysis, read-out probability fit, and Rabi driving respectively. The last column gives the ranges of the hyperfine coupling values from the dipole-dipole calculation (see text).
    }
\end{table}

We use the point-dipole approximation to estimate the hyperfine coupling between the Er$^{3+}$ ion and its neighbouring nuclear spins. The Hamiltonian for this system is

 \begin{equation}
\begin{split}
    H =&\: \omega_S\cdot S_z' + \omega_I \cdot I_z + H_{dd} \\
    H_{dd} = \frac{\mu_0}{4\pi r^{3}} &\left[\bar{\mu}_S\cdot\bar{\mu}_I\ -3r^{-2} (\bar{\mu}_S\cdot\bar{r}) (\bar{\mu}_I\cdot\bar{r}) \right],
\end{split}
\end{equation}

where $\bar{\mu}_S = \overline{\overline{\gamma}}_{\text{Er}^{3+}} \cdot \bar{S}$ and $\bar{\mu}_I = \gamma_{W} \cdot \bar{I}$ are the magnetic moments of the electron and the nuclear spin and $\bar{r}$ is the vector separating the two magnets. Note that the electron spin operators $\bar{S}$ of the dipole-dipole Hamiltonian are not the same compared to the spin operator of the Zeeman term $S_z'$ due to the anisotropy of the gyromagnetic tensor $\overline{\overline{\gamma}}_{\text{Er}^{3+}}$. The dipolar perturbation can be expanded in terms of ($S_x$, $S_y$, $S_z$) and ($I_x$, $I_y$, $I_z$). Since the first term of the Hamiltonian is much larger than the rest, the terms containing $S_x$ and $S_y$ can be ignored, which is known as the secular approximation. The resulting Hamiltonian is

\begin{equation}
    H =\: \omega_S\cdot S_z' + \omega_I \cdot I_z + A S_z'I_z + B S_z'I_x.
\end{equation}

The values for $A$ and $B$ depend non-trivially on the orientation of $B_0$. These terms were calculated numerically for the 10 tungsten sites in the unit cell centered around the paramagnetic impurity. The magnetic field is applied in the $y-z$ plane, with the angle between the field and the z-axis $\theta$ <1º. An angle $\beta = 0.8\pm0.1$º was measured between the c-axis and the $y-z$ plane, which is taken into consideration when computing the hyperfine dipolar interaction. Figure \ref{fig:hyperfine} plots the calculated hyperfine parameters $A$ and $B$ as a function of $\theta$. Table \ref{tab:hyperfine_methods} presents the measured hyperfine parameters for the different ions discussed in the main text, as well as the calculated range for the parameters for Type I, II and III. The table is organized as follows, all ions are sorted depending on their Type assignment (see below) and the last column shows the calculated values from a pure dipole-dipole interaction. We only show the minimum and maximum (min -- max) values for $A$ and $B$ in the range of $\theta$ that is specified. If the minimum and maximum values are closer than 1~kHz, only the average is given. For $\mathrm{Er}_1 - \mathrm{Er}_4$, $|B|$ is calculated from the cross-relaxation method (see App.\ref{section:traces}). For $\mathrm{Er}_5$, $|B|$ is obtained from the Rabi sideband method (App.\ref{section:rabi}).

For the nuclear spins coupled to $\mathrm{Er}_2$ and $\mathrm{Er}_4^{(2)}$, the measured values of $A$ and $B$ are close to those expected from Type III sites, making this site assignment likely. The other nuclear spins have larger values of $|A|$ and $|B|$, implying that they are either Type I or Type II sites; however, the match is not quantitative with any of the sites. Several reasons could explain this. First, local deformations of the crystal could be caused by the larger positive charge of $\mathrm{Er}^{3+}$ compared to the original $\mathrm{Ca}^{2+}$. Second, the measured ions are significantly strained, as evidenced by the fact that they are found several mT above the center of the ensemble line~\cite{wang_single-electron_2023}. In the case of $\mathrm{Er}_5$, the value of $|B|$ measured is significantly larger than the range predicted both for Type I and Type II; however, for larger angles, Type I spins reach such large $B$ values (see Fig.~\ref{fig:hyperfine}), motivating a tentative assignment of $\mathrm{Er}_5$ to a Type I site. We finally note that site assignment should be made easier  by the measurement of the angular dependence of $A$ and $B$.


\section{Electron spin spectroscopy and control}
\label{section:electron}

\begin{figure}[t]
    \centering
    \includegraphics[width=0.9\columnwidth]{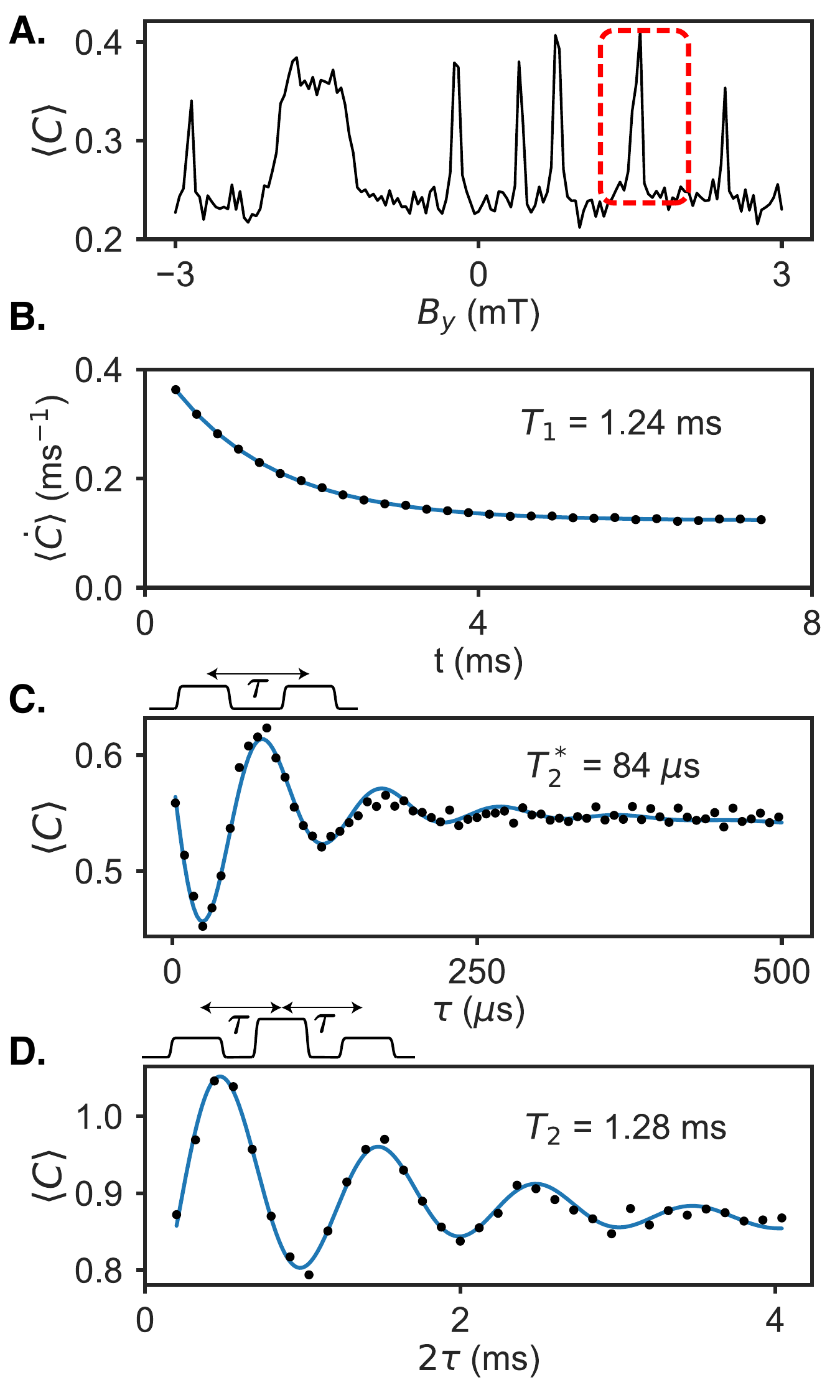}
    \caption{
        \textbf{Electron spin characterization}.
        \textbf{A.} Ensemble averaged counts as a function of magnetic field $B_y$. For this measurement the $Z$ and $X$ magnets are set to persistent mode and $B_z= 446.2$ mT and $B_x = 0.4$ mT. Each peak corresponds to a single Er$^{3+}$ ion. $\mathrm{Er}_5$ is highlighted in red.
        \textbf{B.} Electron spin fluorescence. Count rate after exciting the spin with a $\pi$-pulse. The signal decays with $T_1 = 1.24$ ms to a flat background from the dark counts of the detector.
        \textbf{C.} Ramsey measurement. The coherence of the electron spin is $T_2^* = 81$ $\mu$s. The measurement introduces an arbitrary 1 kHz detuning to the recovery pulse of the sequence for ease of fitting
        \textbf{D.} Echo measurement. The coherence of the electron spin is increased to $T_2$=1.28 ms after one dynamical decoupling step. 
    }
    \label{fig:electron}
\end{figure}

\begin{table}[h]
\begin{tabular}{|l|l|l|l|l|l|}
\hline
                 & \textbf{$\mathrm{Er}_1$} & \textbf{$\mathrm{Er}_2$} & \textbf{$\mathrm{Er}_3$} & \textbf{$\mathrm{Er}_4$} & \textbf{$\mathrm{Er}_5$}  \\ \hline
$T_1$ (ms)       & 2.0$\pm$0.3              & 0.71$\pm$0.1             & 0.90$\pm$0.1         & 0.80$\pm$0.3         & 1.24$\pm$0.1          \\ \hline
$g_0/2\pi$ (kHz) & 3.6$\pm$0.2              & 6.0$\pm$0.2              & 5.3$\pm$0.2          & 5.6$\pm$0.2          & 4.5$\pm$0.1          \\ \hline
$T^*_2$ ($\upmu$s) & -                      & 53$\pm$5                 & -                    & 170$\pm$5            & 81$\pm$2           \\ \hline
$T_2$ (ms)       & -                        & -                        & -                    & 2.05$\pm$0.03        & 1.3$\pm$0.1           \\ \hline

\end{tabular}
\label{tab:electron}
\caption{
    Electron spin parameters for the different ions presented on the main text. Dashes are in place when the measurement was not performed.
}
\end{table}

The electron spin characteristics for $\mathrm{Er}_{1-5}$ are presented in Table \ref{tab:electron}. The values that are not specified were not measured during the experiment. A representative example is $\mathrm{Er}_5$, presented in Fig. \ref{fig:electron}. To perform the initial detection of single spins, the fluorescence signal of the sample is measured after applying a 5 $\upmu$s long square pulse for different values of the magnetic field. When a single spin is in resonance with the superconducting resonator, an increase on the fluorescence signal will be measured after the excitation. This corresponds to every peak in Fig. \ref{fig:electron}A, where $\mathrm{Er}_5$ is highlighted in red. The lifetime of the spin $T_1$ is extracted by measuring directly the fluorescence count rate as a function of time after excitation (Fig. \ref{fig:electron}B). The coupling to the spin resonator $g_0$ is calculated from the Purcell effect, $\Gamma_R = \frac{4g_0^2}{\kappa}$. We note that $g_0$ for Er$_2$ is the highest coupling measured by this technique. Based on the coupling strength we estimate that the ions are within 100~nm of the nanowire (see App.\ref{section:sample}). The coherence times $T_2^*$ and $T_2$ are measured respectively through a Ramsey and Echo sequence (Fig. \ref{fig:electron}C \& D). The different contrast and dark counts between the Ramsey and Echo measurements are due to different operating points of the SMPD.


%

\end{document}